\documentclass{jpsj2}
\usepackage{epsf}

\def\runtitle{Classification of One-Dimensional Quasilattices into MLD Classes}
\def\runauthor{Komajiro \textsc{Niizeki} and Nobuhisa \textsc{Fujita}}

\title
{
Classification of One-Dimensional Quasilattices into Mutual
Local-Derivability Classes
}

\author{
Komajiro \textsc{Niizeki} and
Nobuhisa \textsc{Fujita}$^1$
}

\inst
{
Department of Physics, Graduate School of Science, \\
Tohoku University, Sendai 980-8578\\
$^1$Research Institute of Electrical Communication, \\
Tohoku University, Sendai 980-8577
}

\recdate{\today}

\abst{
One-dimensional quasilattices are classified into mutual local-derivability
(MLD) classes on the basis of geometrical and number-theoretical
considerations. Most quasilattices are ternary, and there exist an infinite
number of MLD classes. Every MLD class has a finite number of quasilattices
with inflation symmetries. We can choose one of them as the representative
of the MLD class, and other members are given as decorations of the
representative. Several MLD classes of particular importance are listed.
The symmetry-preserving decorations rules are investigated extensively.
}

\kword{
quasicrystals, projection method, substitution rule
}

\begin{document}
\sloppy
\maketitle

\section{Introduction}

Quasicrystals have aperiodic ordered structures (for a review, see ref\verb/./ \citen{Ya96})).
The positions of the atoms in a quasicrystal form
a quasilattice (QL), which is a quasiperiodic but discrete set of points.
The classification of QLs is the principal subject of the crystallography
of quasicrystals. There can be various schemes for the classification
depending on the degree of their elaborations. The crudest two are the one based
on the point symmetries and the one on the space
groups. A classification based on mutual local-derivability (MLD) is
more elaborate because two QLs with a common space group can be
inequivalent with respect to MLD (for MLD, see refs\verb/./ \citen{Ba91})
and \citen{Ba99})). A recent report indicates that quasicrystals belonging to
different MLD classes belong to different universality classes with respect
to their one-electron properties. \cite{FuNi00, FuNi01} Therefore, to complete a
classification of QLs into MLD classes is an urgent subject. The purpose of
the present paper is to report a major development on this subject. Our
arguments will be confined to one-dimensional QLs (1DQLs) including those
which appear as 1D sections or 1D projections of octagonal, decagonal and
dodecagonal QLs in 2D and the three types of icosahedral QLs in 3D.
\cite{Ya96, DuKa85}

The subsequent four sections are preparatory: The cut-and-projection method
is introduced in \S \ref{sec:cutproj} (for this method, see refs\verb/./ \citen{Ya96}) and \citen{DuKa85})).
The remaining three preparatory
sections are devoted to arguments on binary QLs.
\cite{LuGoJa93, Lu93, BuFrGaKr98}
That is, their basic properties are summarized in
\S \ref{sec:binary}, and their inflation-symmetry in
\S \ref{sec:infsym}, while their local environments are
investigated in \S \ref{sec:localenv}. Ternary QLs are investigated
in \S \ref{sec:ternary}. The concept of MLD is introduced in
\S \ref{sec:MLD}, and a general theory of the classification of
1DQLs into MLD classes is completed in the same section. We will present in
\S \ref{sec:DR} an algorithm for deriving a decoration rule by
which a ternary QL and its subquasilattice are combined. In the subsequent
section, \S \ref{sec:infs}, we will in the first place introduce
the saw map and then clarify the interrelationship among different QLs in a
single MLD class. The theory is extended in \S \ref{sec:s-MLD} to
the case of the symmetric MLD relationship. In connection with the
gluing condition, we consider in \S \ref{section method} the MLD
relationship from the point of view of the section method. \cite{Ya96}
The final section is devoted to a summary and discussions.

\section{Cut-and-Projection Method}\label{sec:cutproj}

\begin{figure}
\begin{center}\epsfile{file=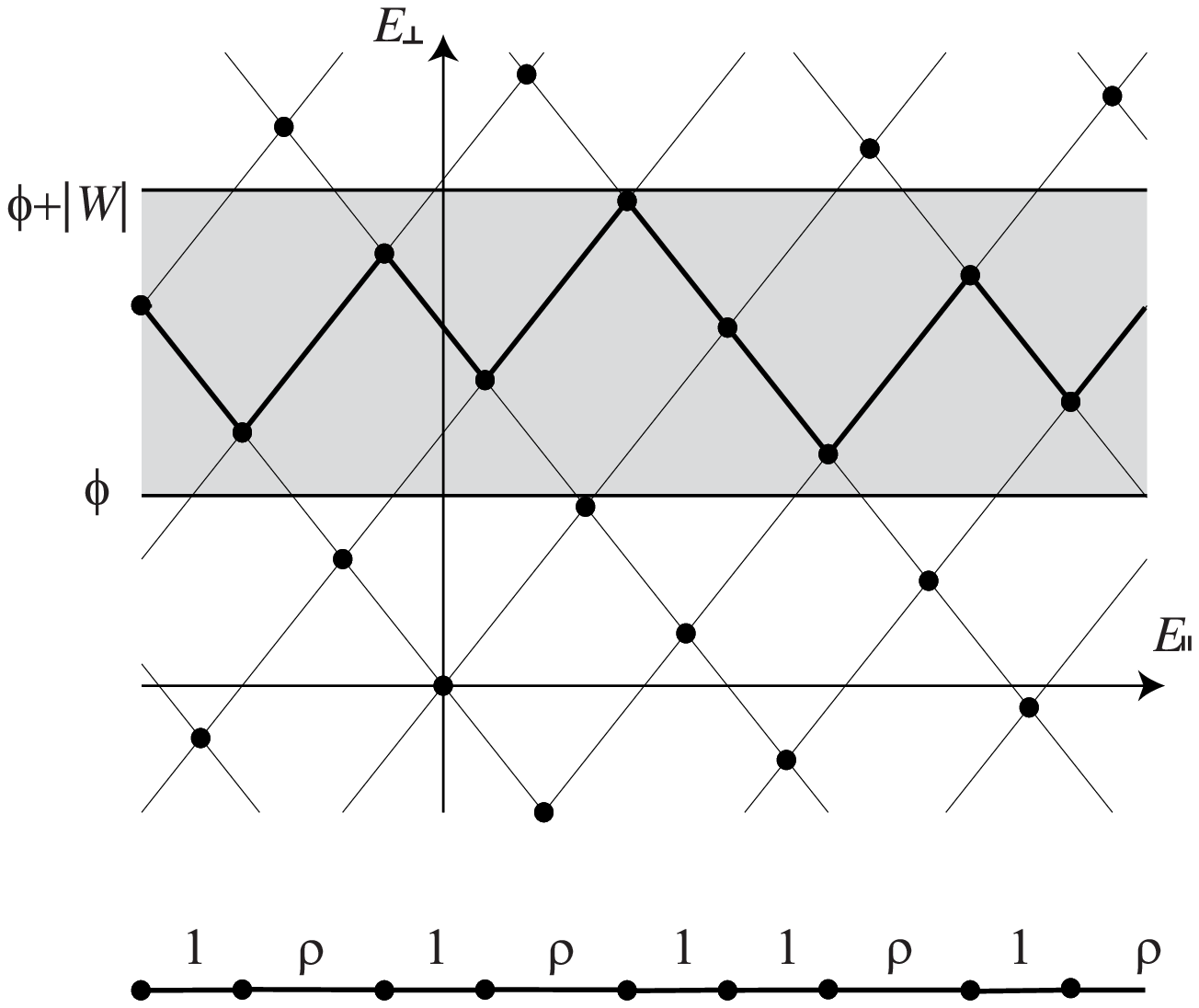,height=10cm}\end{center}
\caption{A derivation of a 1DQL by the cut-and-projection method from a
periodic lattice in 2D. The width of the strip shown by shadowing is chosen
to be equal to the vertical width of a unit cell of the mother lattice, so
that the resulting QL is binary. This example is the $Q_1$ for
$\tau=1+\sqrt{2}\, $ as described in \S 3.}
\label{fig:cut&projection}
\end{figure}

A 1DQL is derived by the cut-and-projection method from a 2D periodic
lattice $\Lambda$, which we shall call a mother lattice. The physical space
$E_\parallel$ contained in the 2D space embedding $\Lambda$ is a line whose
slope is incommensurate with $\Lambda$. The orthogonal complement $E_\perp$
to $E_\parallel$ is called the internal space; thus, $E = E_\parallel
\oplus E_\perp$. Prior to the projection, the mother lattice is cut by a
strip which is parallel to $E_\parallel$, as shown in Fig\verb/./\
\ref{fig:cut&projection}. The strip is specified in terms of its section
$W$ through $E_\perp$. The section is called a window, which is an interval
in $E_\perp$: $W \subset E_\perp$. Let $\phi$ be the position of the left edge of $W$ in
$E_\perp$. Then, we may write $W = \;]\phi, \; \phi + |W|]$ with $|W|$
being the size of $W$. A QL derived from $\Lambda$ is completely
characterized by $W$, and is denoted by $Q(W)$. The parameter $\phi$ is
called a phase of the QL.

The scale of the internal space, $E_\perp$, is irrelevant on the
construction of a QL because the projection in the cut-and-projection
method is made along $E_\perp$. Hence, two mother lattices which differ
only in the scale of the internal space are considered to be isomorphic,
and are usually identified. Note that the mirror operation which fixes
$E_\parallel$ is irrelevant as well.

Let $\Lambda_\parallel$ (resp\verb/./ $\Lambda_\perp$) be the projection of
$\Lambda$ onto $E_\parallel$ (resp\verb/./ $E_\perp$). Then, it is a dense
set in $E_\parallel$ (resp\verb/./ $E_\perp$) because $E_\parallel$ is
incommensurate with $\Lambda$. Therefore, a QL is a discrete subset of
$\Lambda_\parallel$. The density of the lattice-points in $Q(W)$ is given
by $|W|/A_0$ with $A_0$ being the area of a unit cell of $\Lambda$. Two
QLs, $Q(W)$ and $Q(W')$, are known to be locally isomorphic (LI) with each
other if and only if $|W| = |W'|$; locally isomorphic QLs cannot be distinguished
macroscopically from one another. The set of all the QLs having windows
with a specified size form an LI-class; different members in the LI-class
are distinguished by their phases. The mother lattice $\Lambda$ has the
inversion symmetry, and the inversion of $Q(W)$ is given by $Q(-W)$.
\cite{FootNote2}
Hence we can consider every LI-class to have the inversion symmetry. If
$W'$ is another window which is a subset of $W$, $Q(W')$ is a subset of
$Q(W)$. In the subsequent development of the theory, we will usually take
different windows with a common size to be the same window.

A QL divides $E_\parallel$ into intervals, and it gives a 1D tiling of
$E_\parallel$. We shall identify hereafter a QL with the relevant tiling.
There appear three intervals for a generic window, so that a generic QL is
ternary. A binary QL is only obtained for several specific windows to be
given in the subsequent section.

As a consequence of the noncrystallographic point symmetries of
quasicrystals, four quadratic irrationals $\tau_{\rm G} := \frac{1}{2}(1 +
\sqrt{5}\, )$ (the golden mean), $1 + \sqrt{2}\, $ (the silver mean), $2 +
\sqrt{5}\, $ and $2 + \sqrt{3}\, $ accompany the quasicrystals. Let $\tau$
be one of them. Then it is equal to the ratio of two incommensurate periods
along a crystal axis of the relevant quasicrystal. They are algebraic
integers satisfying the quadratic equation $\tau^2 - m\tau + e = 0$ with
$m$ being a natural number and $e = 1$ or $e = -1$. More specifically, $e =
-1$ and $m = 1, \;2$ or 4 for the first three quadratic irrationals above,
while $e = 1$ and $m = 4$ for the last. It follows that the inverse
$\tau^{-1}$ is also an algebraic integer. Let $\bar{\tau}$ be the algebraic
conjugate of $\tau$. Then we obtain $\bar{\tau} = m - \tau$ and
$\tau\bar{\tau} = e$. Our theory will be developed mainly for the case $e =
-1$. The case $e = 1$ can be treated as a modification of the former case.

If $E_\parallel$ and $E_\perp$ are scaled appropriately,
$\Lambda_\parallel$ and $\Lambda_\perp$ turn both into the Z-module ${\bf
Z}[\tau] := \{n_1 + n_2\tau \, |\, n_1, n_2 \in {\bf Z}\}$;
quadratic irrational $\tau$ is the most important parameter
characterizing $\Lambda$. It should be emphasized that the mother lattice
characterized by $2 + \sqrt{5}\, $ is not isomorphic to the one by
$\tau_{\rm G}$. Since $\tau^{-1}$ belongs to ${\bf Z}[\tau]$, ${\bf
Z}[\tau]$ is invariant against the multiplication of $\tau$; $\tau {\bf
Z}[\tau] = {\bf Z}[\tau]$. It follows that two numbers $\tau^n$ and
$\tau^{n+1}$ can generate ${\bf Z}[\tau]$ for any $n \in {\bf Z}$. It is
important in a later argument that two numbers 1 and $p\tau + q$ with $p,
\;q \in {\bf Z}$ can generate ${\bf Z}[\tau]$ only if $p = 1$ or $p = -1$.

We will use several results in the number theory of quadratic fields and
they are summarized in Appendix \ref{NumberTheory}. Several symbols and
technical terms defined in the appendix will be used from now on.

The multiplication of $\tau$ onto ${\bf Z}[\tau]$ induces a linear
automorphism of ${\bf Z}[\tau]$. In particular, the generators are
transformed as
\begin{equation}
\tau(1, \; \tau) = (1, \; \tau)M_0, \label{eqn:mld2}
\end{equation}
with
\begin{equation}
 M_0 = \left( \begin{array}{rr}
        0 & 1  \\
        1 & m
\end{array} \right), \label{eqn:mld3}
\end{equation}
which is unimodular, i.e., ${\rm det}\, M_0 = -1$. It follows from
eq\verb/./(\ref{eqn:mld2}) that $\tau$ is an eigenvalue of $M_0$ and $(1,
\; \tau)$ the relevant left eigenvector. Since $M_0$ is an integer matrix,
the second eigenvalue is given by $\bar{\tau}$ and the second left
eigenvector by $c(1, \; \bar{\tau})$ with $c$ being an indeterminate
factor. Using these results, we can conclude that a regular 2D matrix
defined by
\begin{equation}
 A = \left( \begin{array}{rr}
        1 & \tau  \\
        -1 & -\bar{\tau}
\end{array} \right) \label{eqn:mld4}
\end{equation}
satisfies the equation
\begin{equation}
TA = AM_0, \label{eqn:mld5}
\end{equation}
where $T := \left({\tau \atop 0}\; {0 \atop \bar{\tau}}\right)$ is a
diagonal 2D matrix. Therefore, $M_0$ is diagonalized by $A$ as $AM_0A^{-1}
= T$. The Cayley-Hamilton theorem verifies the equality, $M_0^2 - mM_0 - I
= 0$.

The matrix $A$ is divided into two column vectors ${\bf a}_1$ and ${\bf
a}_2$ so that $A = ({\bf a}_1, \; {\bf a}_2)$. The two column vectors
generate the mother lattice of QLs, $\Lambda := \{n_1{\bf a}_1 + n_2{\bf
a}_2 \, |\, n_1, n_2 \in {\bf Z}\}$. \cite{FootNote3}
The area of a unit cell is given by $A_0 = {\rm det}\, A = \tau +
\tau^{-1}$. Since $M_0$ is unimodular, we can conclude from
eq\verb/./(\ref{eqn:mld5}) that the linear transformation $T$ leaves
$\Lambda$ invariant, so that the transformation is area-preserving. $T$
enlarges the physical space by $\tau$ but shrinks the internal space by
$\tau^{-1} = |\bar{\tau}|$; we shall name the transformation $T$
hyperscaling (for the scaling of QLs, see ref\verb/./\citen{Ni89a})) and
references cited therein). The hyperscaling symmetry of $\Lambda$ gives
rise to a self-similarity (inflation-symmetry) of QLs derived from
$\Lambda$.

Using invariance of $\Lambda$ against the hyperscaling, we can show that
$\tau Q(W) \simeq Q(\tau^{-1}W)$, where the symbol ``$\simeq$" stands for
the LI-relationship. More generally, we obtain $\tau^n Q(W) \simeq Q(W')$
with $W' := \tau^{-n}W$ for any $n \in {\bf Z}$. We shall say that two QLs
with this relation are mutually scaling-LI (SLI). Two QLs which are
mutually SLI are locally isomorphic apart from a difference in the scale by
a factor of the form $\tau^n$. Thus, we can confine our argument to windows
whose sizes belong to the interval $F := \;]1, \;\tau]$, which we shall
call the fundamental interval. It is important to notice that $F$ is not a
subset of $E_\perp$ but is considered to be a subset of the half line
$\;]0, \; \infty[\;$ which is the space of the window size. There exist an
infinite number of SLI-classes of QLs derived from a single mother lattice
and the set of all the SLI-classes has a one-to-one correspondence with the
continuous set, $F$. The classification of QLs with respect to the SLI
relationship is, however, too elaborate in comparison with that with
respect to the MLD relationship.

The mother lattice $\Lambda$ has four types of special points, which are
the points of inversion symmetry in the 2D space embedding $\Lambda$ (for
special points, see ref\verb/./\citen{Ni89d})). Specifically, a special point
is translationally equivalent to 0, $\frac{1}{2}{\bf a}_1$,
$\frac{1}{2}{\bf a}_2$ or $\frac{1}{2}({\bf a}_1 + {\bf a}_2)$, which are
rational points with respect to $\Lambda$. We shall represent the four
types by $\Gamma$, $X$, $Y$ and $C$, respectively, because their
projection onto the real space yield the corresponding four scaled residue
classes introduced in Appendix \ref{scaled residue class}. The hyperscaling
is commutable with the inversion operation, and it induces a similar
permutation among the four types of special points to that among the four
scaled residue classes.

\section{Binary QLs}\label{sec:binary}

The parallelogram spanned by the basis vectors ${\bf a}_1$ and ${\bf a}_2$
is a unit cell of $\Lambda$. The case where the window size is equal to the
width, $1 + \tau^{-1}$, of the unit cell along $E_\perp$ is most important;
$Q_0 := Q(W_0)$ with $|W_0| = 1 + \tau^{-1}$ is a binary QL with the
intervals 1 and $\tau$. The Fibonacci lattice is a special case of $Q_0$
with $\tau = \tau_{\rm G}$, while $Q_0$ gives generalized Fibonacci
lattices for other cases, where $m \ge 2$.

\begin{figure}
\begin{center}\epsfile{file=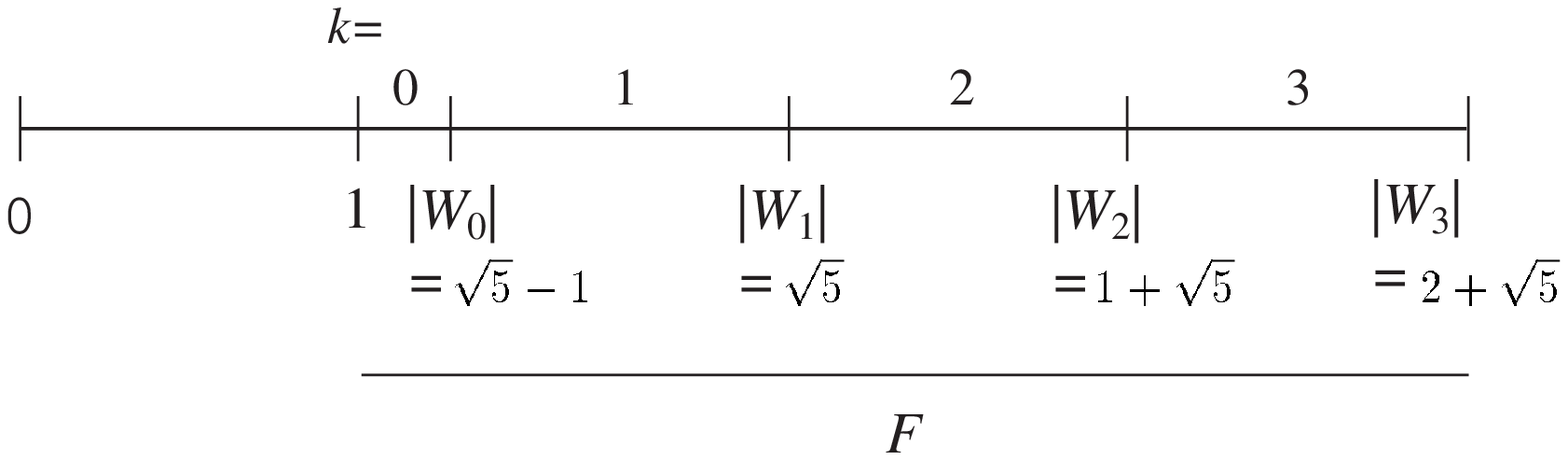,height=3.5cm}\end{center}
\caption{The division of the fundamental interval, $F$, for
$\tau=2+\sqrt{5}\, $ into four classes. The boundaries between different
classes as well as the upper limit of $F$ correspond to the window
sizes of the four binary QLs.}
\label{fig:ClassNumber}
\end{figure}

We present here general results on binary QLs; they will be proven in a
later section. Let us introduce window $W_k$ so that $|W_k| = k + 1 +
\tau^{-1}$ with $k$ being an integer satisfying the inequality $0 \le k \le
m$. Then, we can show that the QL $Q_k := Q(W_k)$ is a binary tiling having
two types of intervals 1 and $\rho$ with $\rho := \tau - k$. The binary
QLs, $Q_0$, $Q_1, \cdots, \; Q_{m-1}$, are not SLI with one another but
$Q_m$ is SLI with $Q_0$ because $|W_m| = \tau |W_0|$. Therefore, we assume
hereafter that the parameter $k$ is an integer variable running on $\{0,
\;1, \;\cdots, \; m -1\}$ unless stated otherwise. A QL obtained from
$\Lambda$ is binary only when it is LI or SLI with one of these QLs, so
that the number of different SLI-classes of binary QLs is equal to $m$.
Remember that the ratio of the two intervals of a binary QL derived from a
single mother lattice takes one of $m$ values of the form $\tau - k$; the
ratio is always larger than 1. $|W_k|$'s are located
equidistantly in the fundamental interval $F$, and $F$ is divided into $m$
subintervals as shown in Fig\verb/./\ \ref{fig:ClassNumber} because
$|W_{m-1}|\;(= \tau)$ coincides with the right edge of $F$.
\cite{FootNote4} We shall call the subintervals classes, each of which is
specified by ``the class number" $k$. The window size $|W_k|$ of a binary
QL, $Q_k$, is located at the right edge of the $k$-th class. The lengths of
classes are equal to 1 except for the zeroth class whose length is equal to
$\tau^{-1}$; the zeroth class is exceptional in many cases to appear later
on. Let us denote by $\Gamma$ the zeroth class. Then the fundamental interval
is divided as $F = \Gamma \cup \Gamma^{\rm c}$ with $\Gamma^{\rm c}$ being
the complement of $\Gamma$ in $F$. Remember that the boundary between
$\Gamma$ and $\Gamma^{\rm c}$ is located at $|W_0| = 1 + \tau^{-1}$.

Since the pair (1, $\rho$) of numbers as well as (1, $\tau$) can generate
${\bf Z}[\tau]$, there exists a unimodular matrix satisfing the equation,
$(1, \; \tau) = (1, \; \rho)K$:
\begin{equation}
 K = \left( \begin{array}{lc}
        1 & k  \\
        0 & 1
\end{array} \right).\label{eqn:mld6}
\end{equation}
Equation (\ref{eqn:mld2}) is transformed with the new generators into
\begin{equation}
\tau(1, \; \rho) = (1, \; \rho)M, \label{eqn:mld7}
\end{equation}
with $M$ being a unimodular matrix,
\begin{equation}
M = KM_0K^{-1}. \label{eqn:mld7b}
\end{equation}
To be more specific,
\begin{equation}
 M = \left( \begin{array}{lc}
        k &  k(m - k) + 1  \\
        1 & m - k
\end{array} \right) =: M_k, \label{eqn:mld8}
\end{equation}
which is consistent with eq\verb/./(\ref{eqn:mld3}) for $k = 0$.
Note that $M$ satisfies the same quadratic equation as that for $M_0$, so
that ${\rm Tr}\, M = m$.

Equation (\ref{eqn:mld7}) reads as follows: scaling up the interval 1 by
$\tau$ yields a new interval which is divided into $k$ intervals of the
type ``1" and one interval of the type ``$\rho$" and similarly goes it for
scaling up the interval $\rho$; $M$ is uniquely determined by
eq\verb/./(\ref{eqn:mld7}) because $\rho$ is irrational.

It is important in a later argument that $M_0$ is represented as $M_0 =
L'L^{m - 1}$ with $L = \left({1 \atop 0}\; {1 \atop 1}\right)$ and $L'
=\left({0 \atop 1}\; {1 \atop 1}\right)$, so that $M_k = L^kL'L^{m - k -
1}$ because $K = L^k$.

It follows from Eqs.(\ref{eqn:mld5}) and (\ref{eqn:mld7b}) that $M$ is
diagonalized by $B := AK^{-1}$, and hence $TB = BM$ or, equivalently,
$BMB^{-1} = T$. The two column vectors ${\bf b}_1$ and ${\bf b}_2$ of $B$
are given by
\begin{equation}
{\bf b}_1 = {\bf a}_1, \; {\bf b}_2 = {\bf a}_2 - k{\bf a}_1, \label{eqn:mld11}
\end{equation}
so that
\begin{equation}
 B = \left( \begin{array}{rr}
        1 & \rho  \\
        -1 & \sigma
\end{array} \right), \label{eqn:mld9}
\end{equation}
where $\sigma := -\bar{\rho} = k + \tau^{-1} = k - m + \tau$. The two
vectors, ${\bf b}_1$ and ${\bf b}_2$, form a new set of basis vectors of
$\Lambda$. The parallelogram spanned by them is a unit cell of $\Lambda$,
and its width along $E_\perp$ is given by $1 + \sigma$, which is equal to
$|W_k|$. Therefore, the binary QL, $Q_k := Q(W_k)$, is written as
\begin{equation}
Q_k = \{n_1 + n_2\rho \, |\, n_1, \;n_2 \in {\bf Z}\; {\rm and} \; -n_1 +
n_2\sigma \in W_k \}, \label{eqn:mld12}
\end{equation}
where $n_1 + n_2\rho$ and $-n_1 + n_2\sigma$ are the projections of the
lattice vector $n_1{\bf b}_1 + n_2{\bf b}_2 \in \Lambda$ onto $E_\parallel$
and $E_\perp$, respectively. We can conclude from this equation that the
ratio of the frequency of occurrence of the interval 1 to that of $\rho$ is
equal to $\sigma$. Equivalently, the components of the column vector $w :=
{}^{\rm t}(\sigma, \; 1)$ are proportional to the frequencies of the two
intervals in the QL. By these reasons, we shall call the row vector $u :=
(1, \; \rho)$ the interval vector, while $w$ the frequency vector. We have
$\sigma < 1$ only for the special case $k = 0$, where $M = M_0$, $\rho =
\tau$ and $\sigma = \tau^{-1}$. It is important in a later argument
whether this $\sigma$ is smaller or larger than $\frac{1}{2}$, that is,
$\tau$ is larger or smaller than 2. The latter is the case only when $m =
1$ ($\tau = \tau_{\rm G}$).

Although two quantities $\rho$ and $\sigma$ depend not only on $\tau$ but also on
the value of the class number $k$, we will not explicitly show it to
avoid a notational complication. We must remember that these quantities
are context dependent. It is also important in a later argument that the
size of the window of every binary QL is a $\tau$-integral which is not
divisable by 2.

The binary nature of the QL, eq\verb/./(\ref{eqn:mld12}), is verified geometrically by
the following observation: there exists a unique zigzag line which passes
all the lattice points in the cut piece of $\Lambda$ by the relevant strip
and, besides, the zigzag line is formed of bonds of the type ${\bf b}_1$ or
${\bf b}_2$ (see Fig\verb/./\ \ref{fig:cut&projection}).

The two row vectors of $B$ are the left eigenvectors of $M$, while the two
column vectors of $B^{-1}$ are the right ones. In particular, the first
column vector of $B^{-1}$ is proportional to the frequency vector $w$, so
that $w$ is also a right eigenvector of $M$. The eigenvalue $\tau$ of $M$
is the Frobenius eigenvalue, while the interval vector $u$ (resp\verb/./
the frequency vector $w$) is the left (resp\verb/./ right) Frobenius
eigenvector.

\section{Inflation-Symmetry}\label{sec:infsym}

We begin with the case of the Fibonacci lattice (FL), which is composed of
two types of intervals, $1$ and $\tau = \tau_{\rm G}$. It is well known
that the FL has an inflation-symmetry represented by the inflation rule
(IR), $1' = \tau$ and $\tau' = 1\tau$; an FL is transformed into another FL
if the intervals in the original FL are substituted as indicated by the IR.
On this substitution, each interval is scaled up by $\tau$.

We shall investigate the IR from the point of view of the
cut-and-projection method. The interval 1 is always isolated in the FL. If
a lattice point in the FL is deleted whenever it is located at the right
end of an interval of the type $1$, the resulting QL is a binary tiling
of the two types of intervals, $1' := \tau$ and $\tau' := 1 + \tau$, which
are the scaled versions of $1$ and $\tau$ by the ratio $\tau$. Let $Q$ be
the original FL and $Q'$ the new one obtained in this way. Then they are
SLI by the inflation-symmetry of FL. More precisely, $\tau Q \simeq Q'$. If
$W$ and $W'$ are the windows of $Q$ and $Q'$, respectively, we find that
$W' \subset W$ from $Q' \subset Q$ and $|W'| = \tau^{-1}|W|$ from $\tau Q
\simeq Q'$. It can be shown that $W'$ is located in $W$ so that the right
ends of the two windows coincide.

The relation between the two sets of intervals is cast into a single
equation: $(1', \; \tau') = (1, \; \tau)M_0$, where $M_0$ is given by
eq\verb/./(\ref{eqn:mld3}) with $m = 1$, i.e., $M_0 = \left({0 \atop 1}\;
{1 \atop 1}\right)$.
Therefore, the IR of the FL is dominated by eq\verb/./(\ref{eqn:mld2}) with
eq\verb/./(\ref{eqn:mld3}). The corresponding equation in the general case
is eq\verb/./(\ref{eqn:mld7}) with eq\verb/./(\ref{eqn:mld8}). It can be
shown that the QL given by eq\verb/./(\ref{eqn:mld12}) has an
inflation-symmetry represented by the IR:
\begin{equation}
1' = 1^k\rho, \quad \rho' = (1^k\rho)^{m - k}1. \label{eqn:mld15}
\end{equation}
For example, the IR for the case of $m = 4$ and $k = 2$ is given by $1' =
11\rho$, $\rho' = 11\rho11\rho1$ with $\tau = 2 + \sqrt{5}\, $ and $\rho =
\sqrt{5}\, $. Since the inflation-symmetry of the relevant QL is dominated
by the matrix $M$, we call it the inflation matrix of the QL.
The majority interval of the two is 1 or $\rho$ if $\sigma > 1$ or $\sigma
< 1$, respectively. In the former (resp\verb/./ latter) case, the interval
$\rho$ (resp\verb/./ 1) is isolated, while the interval 1 (resp\verb/./
$\rho$) appears as $\rho1^k\rho$ or $\rho1^{k + 1}\rho$ (resp\verb/./
$1\rho^m1$ or $1\rho^{m + 1}1$). We shall call $k$ or $m$ the lower
repetition number or the repetition number, for short, of the relevant QL.

We shall digress, for the moment, from the main course to the case $e = 1$.
We may assume that $m \ge 4$ because the relevant mother lattice for the
case of $m = 3$ is isomorphic to that characterized by $\tau_{\rm G}$ on
account of the equality, ${\bf Z}[\tau] = {\bf Z}[\tau_{\rm G}]$. The case
$e = 1$ cannot be incorporated in our formulation above because the sign of
$\tau{\bar \tau} \;(= 1)$ is opposite to that of the case $e = -1$.
Consequently, the unimodular matrix defined by eq\verb/./(\ref{eqn:mld2})
is not a Frobenius matrix. The inflation matrix $M \in GL^+(2, \;{\bf Z})$
of a binary QL whose ratio of the inflation-symmetry is $\tau$ should
satisfy the equation: $M^2 - mM + I = 0$, so that ${\rm Tr}\, M = m$ and
${\rm det}\, M = 1$. Hence, $M$ takes the form:
\begin{equation}
 M = \left( \begin{array}{lc}
        k & k(m - k) - 1  \\
        1 & m - k
\end{array} \right) =: M'_k.\label{eqn:mld15b}
\end{equation}
with $k = 1, \;2, \;\cdots, \; m -1$ (cf. eq\verb/./(\ref{eqn:mld8})); the
case $k = 0$ must be excluded, which is an essential proviso of the present
case of $e = 1$. The matrix above is given, alternatively, by $M'_k = L^{k
- 1}L''L^{m - k - 1}$ with $L'' = \left({1 \atop 1}\; {0 \atop 1}\right)$.
Note that $M'_k \equiv M_k \bmod 2$.

The eigenvalues of $M'_k$ are $\tau$ and ${\bar \tau}$ with ${\bar \tau} =
\tau^{-1}$, and the left Frobenius eigenvector is $(1, \; \rho)$ with $\rho
= \tau - k$; $M$ is diagonalized by the matrix defined by
eq\verb/./(\ref{eqn:mld9}) with $\sigma = k - \tau^{-1}$, which is smaller
than one if $k = 1$. Thus, our formulation above is basically
applicable, with the proviso above, to the present case. In particular,
there exist $m - 1$ SLI classes of binary QLs with inflation symmetries;
their representatives are $Q_k := Q(W_k)$ with $|W_k| = 1 + k - \tau^{-1}$.
The fundamental interval is divided into $m - 1$ classes (subintervals). It
is divided, alternatively, as $F = \Gamma \cup \Gamma^{\rm c}$, where
$\Gamma$ must be taken to be the class 1.

It can be shown that $Q_k$ has the IR:
\begin{equation}
1' = 1^k\rho, \quad \rho' = (1^k\rho)^{m - k - 1}1^{k-1}\rho, \label{eqn:mld15c}
\end{equation}
which is consistent with the inflation matrix eq\verb/./(\ref{eqn:mld15b}).
Note that $\sigma > \frac{1}{2}$ for the case of $k = 1$, which is similar
to the case of $m = 1$ and $e = -1$. It is important in a later section
that the interval $\rho$ can take the value, $1 - \tau^{-1}$, which is shorter
than 1.

Most of the results to be given in later sections will be applicable
directly, or with a minor modification, to the present case of $e = 1$; the proviso
above is essential.

A binary QL is numbered as $Q_k$ by $k$ for each value of $\tau$, so that
we must mind which value of $\tau$ is relevant when a binary QL is referred as $Q_k$.

\section{Local Environments}\label{sec:localenv}

Let us express the number $n_1 + n_2\rho$ appearing in
eq\verb/./(\ref{eqn:mld12}) by $\nu$. Then, $-n_1 + n_2\sigma = -{\bar
\nu}$, which we express as ${\tilde \nu}$: ${\tilde \nu} := -{\bar \nu}$.
Since ${\bf Z}[\rho] = {\bf Z}[\tau]$, we can define a linear map from
${\bf Z}[\tau]$ onto itself by: $\nu \in {\bf Z}[\tau] \to {\tilde \nu} \in
{\bf Z}[\tau]$. The map is a bijection and is recursive: ${\tilde {\tilde
\nu}} = \nu$. Using the map, we can write a QL derived from $\Lambda$ as
\begin{equation}
Q = \{\nu \, |\, \nu \in {\bf Z}[\tau] \;{\rm and} \; 0 \le {\tilde \nu} -
\phi < |W| \}, \label{eqn:mld16}
\end{equation}
where $\phi$ is the phase of the window. Note that the size $|W|$ in this
expression can be generic.

Different sites of $Q$ have different environments, and the environment of
a site $\nu \in Q$ is determined by the position of the number $\phi(\nu)
:= {\tilde \nu} - \phi$ in the interval $\;]0, \;|W|]$; we shall name
$\phi(\nu)$ the local phase of the site $\nu$. If a site is in a local
environment, there exist an infinite number of sites with the same local
environment. A subinterval in $W$ is associated with the local environment,
and the density of such sites is proportional to the length of the
subinterval, which we shall call a subwindow. It must be emphasized that
every local environment has its own subwindow. A site in a local
environment is shown, hereafter, by ``$\bullet$" as $1\bullet\rho$, which
means that the site sits between the interval 1 on the left and $\rho$ on
the right. A subwindow of a specified environment is shown, for example, as
$[1\bullet\rho]$. Owing to the inversion symmetry of an LI-class, a
subwindow for a specified environment and another for the mirror image of
the environment have a common size, and are located symmetrically in the
original window. In particular, a subwindow for a symmetric environment is
located in cocentric with the original window.

The following equality can be readily shown for any $\lambda \in {\bf
Z}[\tau]$:
\begin{equation}
\lambda + Q(W) = Q(-{\tilde \lambda} + W); \label{eqn:mld16x}
\end{equation}
that is, a translation of $Q(W)$ in the real space by $\lambda$ is another
QL whose window is a shift in the internal space of the original window by
$-{\tilde \lambda}$.

If we take two sites $\nu$ and $\nu'$ of $Q$, the relative local phase
between the two sites is given by ${\tilde \lambda}$ with $\lambda := \nu'
- \nu$. As a consequence, we obtain an inequality: $|{\tilde \lambda}| <
|W|$. We can verify from this result the following proposition, which is
important in the subsequent development of our theory:

\begin{quotation}
\noindent {\bf P1}: Let $\lambda \in {\bf Z}[\tau]$ and assume that $0 <
|{\tilde \lambda}| < |W|$. Then, $W$ is divided into two subwindows with
sizes $|{\tilde \lambda}|$ and $|W| - |{\tilde \lambda}|$ so that a site
whose local phase belongs to the second subwindow has a neighbor in the QL
at a distance given by $\lambda$, where the first subwindow is located on
the right or left of the second depending on whether ${\tilde \lambda}$ is
positive or negative, respectively.
\end{quotation}
Let $W_1$ and $W_2$ be the first and the second subwindows, respectively,
of {\bf P1}. Then we can distinguish the two cases of ordering of
the subwindows by the equation $W = W_1 \vee W_2$ or $W = W_2 \vee W_1$;
the ordering must not be changed.

Let $W'$ be a subwindow of $W$ associated with a local environment. Then we
can conclude from {\bf P1} or eq\verb/./(\ref{eqn:mld16x}) that the
distance of the left end of $W'$ from either end of $W$ belongs to ${\bf
Z}[\tau]$; we can assign the symbol ``l" or ``r" for the left end of $W'$
according as the relevant end of $W$ is the left one or the right one,
respectively. The same condition is satisfied by the right end of $W'$ as
well. These conditions strongly restrict the size and the position of $W'$
in $W$. There exist four cases of the combinations of the types of the two
ends of $W'$: ``lr", ``rl", ``ll" and ``rr". The size of $W'$ satisfies
one of the following three equations:
\begin{equation}
     \begin{array}{lll}
       |W'| \equiv |W| & \bmod \;{\bf Z}[\tau] & \mbox{for ``lr"}, \\
       |W'| \equiv -|W| & \bmod \;{\bf Z}[\tau] & \mbox{for ``rl"}, \\
       |W'| \in {\bf Z}[\tau] & {} & \mbox{for ``ll" or ``rr"};
     \end{array} \label{eqn:mld16y}
\end{equation}
the three cases of the equation will be referred to as the positive,
negative or homogeneous case, respectively.

We shall apply {\bf P1} to a binary QL with $\rho = \tau - k$. Using
the equations ${\tilde 1} = -1$ and ${\tilde \rho} = \sigma$, we can
reconfirm that a window yields a binary QL which has two types of intervals
with the lengths 1 and $\rho$ as long as $|W| = 1 + \sigma$; this window is
divided into two subwindows as $W = [*\bullet\rho] \vee [*\bullet 1]$ with
$|[*\bullet 1]| = \sigma$ and $|[*\bullet\rho]| = 1$, where ``$*$" stands
for an arbitrary interval. The mirror image of this division is $W =
[1\bullet *] \vee [\rho\bullet *]$. These divisions are consistent with
that the ratio of the frequency of the interval 1 to that for $\rho$ is
equal to $\sigma$. We shall distinguish the two divisions by calling them
the R-division or the L-division, respectively; the two are the mirror
images of each other.

We next proceed to division of a window with respect to local environments
of the type $\alpha\bullet\beta$ with $\alpha$ and $\beta$ being some
specified intervals. We shall call this division an LR-division, which is
symmetric. The LR-division is performed by using the results for the
R-division and the L-division; an LR-division has two division points,
which are derived from those of the R-division and the L-division. The
result is different according as $\sigma$ is larger or smaller than 1. In
the former case, $W = [1\bullet\rho] \vee [1\bullet1] \vee [\rho\bullet
1]$; the sizes of these subwindows are 1, $\sigma - 1$ and 1, respectively,
which is consistent with the identity: $[1\bullet1] \vee [\rho\bullet 1] =
[*\bullet 1]$ and $[1\bullet\rho] \vee [1\bullet1] = [1\bullet *]$. The
subwindow of the type $[\rho\bullet\rho]$ is absent in this LR-division,
which is consistent with that the interval $\rho$ is isolated in the QL but
the interval $1$ can repeat. On the other hand, the window is divided as $W
= [1\bullet\rho] \vee [\rho\bullet\rho] \vee [\rho\bullet 1]$ for $\sigma <
1$; the sizes of these subwindows are $\sigma$, $1 - \sigma$ and $\sigma$,
respectively.

\section{Ternary QLs}\label{sec:ternary}

We begin with a binary QL for which $\sigma > 1$. We consider what happens
when the left side of its window is cut by the amount $\delta$ with $0 <
\delta < 1$. A part of the lattice points of the original QL disappear
then; these lattice points are in the local environment of the type
$[1\bullet\rho] $. Consequently, the longer intervals of the length $\omega
:= 1 + \rho$ appear; the new QL is a ternary QL composed of the three types
of intervals 1, $\rho$ and $\omega$. The fact that the length of the third
interval is the sum of those of the first two is one of most important
features of present ternary QLs, and will be referred to as the addition
rule. Let $W$ with $|W| = 1 + \sigma - \delta$ be the window of the ternary QL.
Then, its R-division is obtained in a similar way to the case of a binary
QL, yielding $W = [*\bullet \rho] \vee [*\bullet\omega] \vee [*\bullet1]$
with $|[*\bullet \rho]| = 1 - \delta$, $|[*\bullet\omega]| = \delta$ and
$|[*\bullet 1]| = \sigma - \delta$; the frequencies of 1 and $\rho$ have
lowered by the amount $\delta$ because a part of the intervals of these
types are used to form new intervals of the type $\omega$. In the limiting
case where $\delta = 1$, we obtain a binary QL with two types of intervals,
1 and $\omega$. On the other hand, the L division is the mirror image of
the R division. In conclusion, the interval vector and the corresponding
frequency vector of the ternary QL are given, respectively, by
\begin{equation}
u = (1, \;\rho, \;\omega), \quad w = {}^{\rm t}(\sigma - \delta, \;1 -
\delta, \;\delta), \label{eqn:mld16b}
\end{equation}
where the frequency vector is normalized so that its components are equal
to the sizes of the relevant subwindows. The expression for the frequency
vector shows that the interval 1 remains as the majority for $0 < \delta <
1$ as long as $\sigma > 2$. The situation is different if $1 < \sigma < 2$;
the majority interval changes from $1$ to $\omega$ as $\delta$ is increased
from 0 to 1.

In the case $\sigma < 1$, $\delta$ must satisfy the inequality $0 < \delta
< \sigma$. The window satisfies the inequality $1 < |W| < |W_0|$, and the
limiting case $|W| = 1$ realized when $\delta = \sigma$ yields a binary QL
with the intervals $\rho = \tau$ and $\omega = 1 + \tau$; this binary QL is
LI with $\tau Q_{m-1}$. It is important in a later argument that $|W| =
2\sigma$ when $\frac{1}{2} < \sigma < 1$ and $\delta = 1 - \sigma$.

The two parameters, $\rho = \tau - k$ and $\sigma = k + 1/\tau$, are
determined by the class number $k$ of the class to which $|W|$ belongs, and
so are $u$, $\delta$ and $w$; $k$ is the maximal integer which does not
exceed $|W| - 1/\tau$. Therefore, we can assign the class number $k$ to the
ternary QL as well. The two functions, $\sigma$ and $k$, of $|W|$ are
piece-wise flat, and have a finite number of discontinuities at $|W| =
|W_k|$.

The frequency of one of the three intervals vanishes in the binary limit,
where $|W|$ tends to one of the two boundaries of the relevant class. The
repetition number is defined for a ternary QL as well but it will be
investigated later on.

\begin{figure}
\begin{center}\epsfile{file=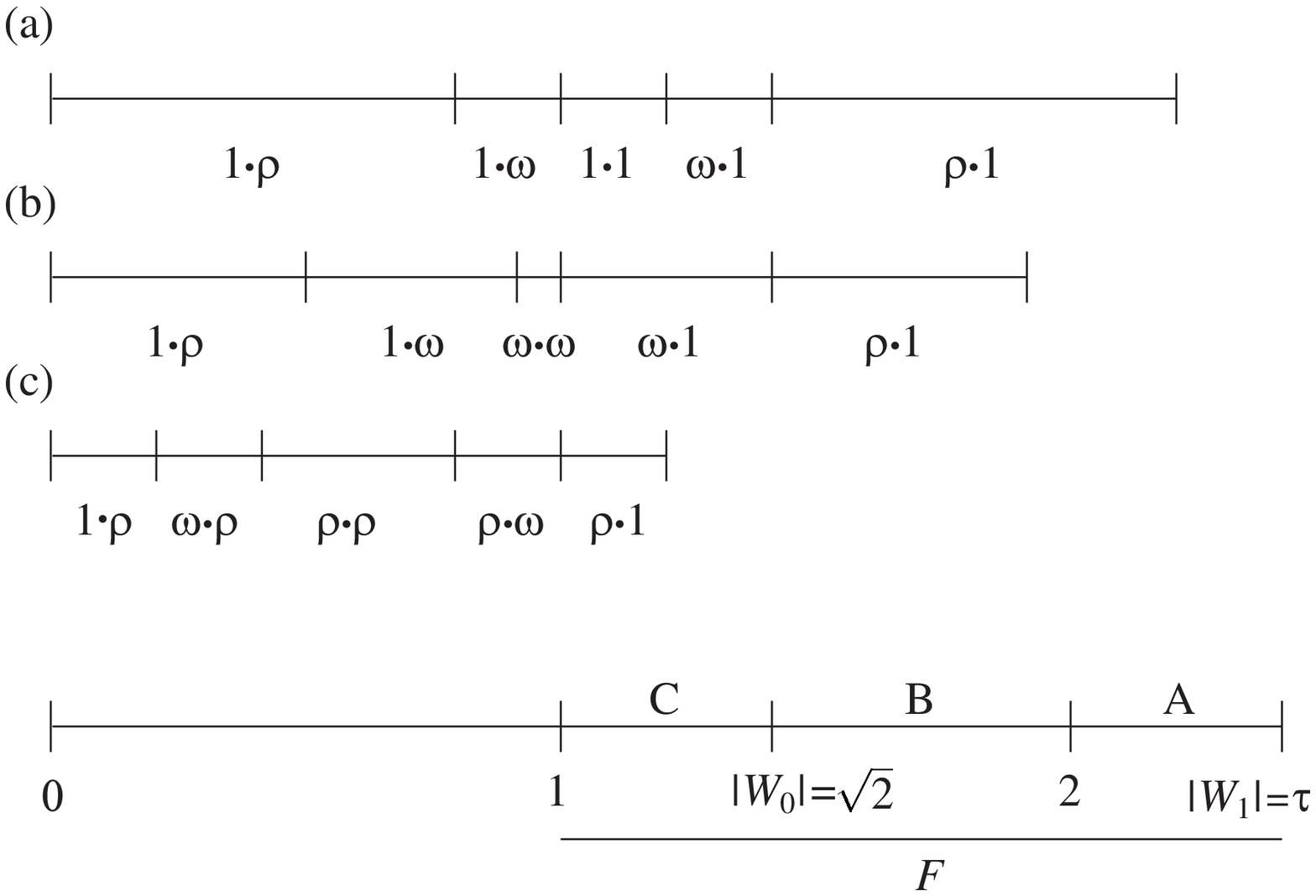,height=8cm}\end{center}
\caption{The LR-divisions of three windows, which yield ternary QLs with
$\tau=1+\sqrt{2}\, $. The LR-division of a window depends on which of the
subintervals, A, B and C, the window size, $|W|$, belongs to. The window
sizes are $|W|=\frac{1}{2}(3+\sqrt{2}\, )\in A$ for (a),
$|W|=\frac{1}{2}+\sqrt{2}\, \in B$ for (b) and
$|W|=\frac{1}{2}(1+\sqrt{2}\, )\in C$ for (c). At the bottom is the division
of the fundamental interval.}
\label{fig:LR-Divisions}
\end{figure}

The LR-division of the window of a ternary QL is performed in a similar way
as the case of binary QLs by using the results for the R-division and the
L-division. The result depends on the window size. Prior to present the
result, we divide $\Gamma^{\rm c}$, the second part of the fundamental
interval $F = \;]1, \;\tau]$, into two subintervals $A$ and $B$ at $|W| =
2$, namely, $\Gamma^{\rm c} = B \cup A$, so that $F = C \cup B \cup A$,
where
\begin{equation}
A := \;]2, \;\tau], \quad B := \;]|W_0|, \;2], \quad C := \;]1, \;|W_0|].
\label{eqn:mld16c}
\end{equation}
Then the LR-division assumes:
\begin{equation}
     \begin{array}{ll}
       W = [1\bullet\rho] \vee [1\bullet\omega] \vee[1\bullet1] \vee
[\omega\bullet 1] \vee  [\rho\bullet 1], & |W| \in A, \\
       W = [1\bullet\rho] \vee [1\bullet\omega] \vee [\omega\bullet\omega]
\vee [\omega\bullet 1] \vee [\rho\bullet 1], & |W| \in B, \\
       W =  [1\bullet\rho] \vee [\omega\bullet\rho] \vee [\rho\bullet\rho]
\vee [\rho\bullet\omega] \vee [\rho\bullet 1], & |W| \in C,
     \end{array} \label{eqn:mld17}
\end{equation}
which are symmetric. These divisions are illustrated in Fig\verb/./\
\ref{fig:LR-Divisions}. The sizes of the five subwindows are given for the
three cases as:
\begin{equation}
     \begin{array}{ccccc}
       1 - \delta & \delta & \sigma - 1 - \delta & \delta & 1 - \delta, \\
       1 - \delta & \sigma - 1 & 1 - \sigma + \delta & \sigma - 1 & 1 -
\delta, \\
       \sigma - \delta & \delta & 1 - \sigma - \delta & \delta & \sigma -
\delta,
     \end{array} \label{eqn:mld18}
\end{equation}
respectively. It is observed in eq\verb/./(\ref{eqn:mld17}) that only one
of the three types of intervals can repeat in a ternary QL but it is not
necessary the majority interval. Note that the subwindow of the type
$[x\bullet x]$ occupies a central region of $W$.

The case $|W| = 2$ is in a sense singular because $|W|$ is located at the
boundary between the two intervals $A$ and $B$; the central subwindow
($[1\bullet1]$ or $ [\omega\bullet\omega] $) vanishes:
\begin{equation}
     \begin{array}{ll}
        W = [1\bullet\rho] \vee [1\bullet\omega] \vee [\omega\bullet 1]
\vee [\rho\bullet 1].
     \end{array} \label{eqn:mld17p}
\end{equation}
We obtain $\rho = \tau - 1$, $\sigma = 1 + \tau^{-1}$ and $\delta = \sigma
- 1$ ($= \tau^{-1}$) for this case. We shall denote the relevant QL as
$Q_{\rm b}$, where the suffix ``b" stands for ``boundary". It follows from
the above equation that no intervals of the three types repeat in $Q_{\rm
b}$. More precisely, $Q_{\rm b}$ is obtained from a binary QL composed of
$\rho$ and $\sigma$ by inserting 1 into every junction of the binary QL, so that
$Q_{\rm b}$ is virtually binary. The identity of the binary QL will be
revealed in \S \ref{sec:DR}.

We have assumed tacitly that $\sigma < \frac{1}{2}$ for the case C. The
situation is different for the case of $\sigma > \frac{1}{2}$, which is
realized when $\tau=\tau_{\rm G}$; the value of $\sigma$ is bound in this
case to $1/\tau$. The interval $C$ should be replaced in this case by
$\;]2\sigma, \;1 + \sigma]$, and the remaining interval, $D := \;]1,
\;2\sigma]$, has a different LR-division as given by
\begin{equation}
       W =  [1\bullet\rho] \vee [\omega\bullet\rho] \vee
[\omega\bullet\omega] \vee [\rho\bullet\omega] \vee [\rho\bullet 1], \;|W|
\in D = \;]1, \;2\sigma]. \label{eqn:mld17b}
\end{equation}
The sizes of the five subwindows are
\begin{equation}
     \begin{array}{ccccc}
        \sigma - \delta & 1 - \sigma & \sigma - 1 + \delta & 1 - \sigma &
\sigma - \delta
     \end{array} \label{eqn:mld18b}
\end{equation}
with $\delta = 1 + \sigma - |W|$. The second type of the boundary QLs,
$Q'_{\rm b}$, is obtained for the window with $|W| = 2\sigma$.
It has similar properties to those of $Q_{\rm b}$. Note that $Q'_{\rm b}$
is only available for $m = 1$, while $Q_{\rm b}$ for $m > 1$.

It is evident that the quantity $|W_0|$ appeared in this section should be
replaced by $|W_1| = 2 - 1/\tau$ for $e = 1$. Also, a similar modification
to the one in the preceding paragraph is necessary for the case $C$ if
$e = 1$ because $\sigma = 1 - 1/\tau > \frac{1}{2}$ for this case. In order
to avoid such complications we shall develop our theory mainly for the
cases of $e = -1$ and $m \ge 2$. Remember that both $Q_{\rm b}$ and
$Q'_{\rm b}$ are available for $e = 1$.

It is important in a later section that the types of the six boundaries of
the LR-division form the sequence ``lrlrlr" for the cases A and C but
``lrrllr" for the cases B and D; these sequences are mirror symmetric if
``l" and ``r" are considered to be the mirror images of each other. Also it
is important in a later section that the windows of the boundary QLs,
$Q_{\rm b}$ and $Q'_{\rm b}$, are even $\tau$-integrals.

A ternary QL includes an infinite variation of sequences with the mirror
symmetry. Their centers of the symmetry are the projections of the special
points of the mother lattice \cite{Ni89d}, so that they can be classified
into four types, $\Gamma$, $X$, $Y$ and $C$, which correspond to the four
residue classes included in $\frac{1}{2}{\bf Z}[\tau]$. A center of the
type $\Gamma$ is a lattice point of the QL, while those of other three
types are located at the centers of the three types of intervals. More
precisely, a center of the type $X$ is located at the center of the
interval 1, while those of $Y$ and $C$ (resp\verb/./ $C$ and $Y$) are at
the centers of the intervals, $\rho$ and $\omega$, respectively, if the
class number of $\rho := \tau - k$ is even (resp\verb/./ odd). A QL has a
global mirror symmetry if the center of the window of the QL coincides with
a point which is a projection of a special point of the mother lattice onto
the internal space. Hence, there exists only four types of QLs with exact
mirror symmetries, and are specified by the four symbols, $\Gamma$, $X$,
$Y$ and $C$.

\section{Classification into Mutual Local-Derivability Classes}\label{sec:MLD}

If two QLs, $Q$ and $Q'$, are locally derivable from each other, we shall
say that the two QLs are in the relation to be called mutual
local-derivability (MLD), where the two QLs are considered to be different
subsets of a common 1D space, i.e., the physical space. Since the MLD
relationship is transitive, all the QLs derived from a single mother
lattice are classified with respect to this relationship. \cite{Ba91} It
should be emphasized that two QLs can be MLD only when their mother
lattices are characterized by a common quadratic integer.

It is important to distinguish strong MLD (SMLD) and weak one. A definition
of SMLD is reduced to that of strong local-derivability; a QL, $Q$, is
strongly locally derivable from another QL, $Q'$, if i) $Q$ is locally
derivable from $Q'$ and, besides, ii), for every lattice point of $Q$, the
distance between it and the relevant local structure of $Q'$ is bounded. On
the other hand, the weak MLD is defined as follows: two QLs, $Q$ and $Q'$,
are weakly MLD \cite{FootNote9} if $Q$ and $\tau^n Q'$ are SMLD for some $n
\in {\bf Z}$. By definition, two QLs, $Q$ and $Q'$, are weakly MLD if they
are SMLD. A weak MLD can be called simply as MLD. Note that the original
MLD defined by M. Baake et al \cite{Ba91} is SMLD in our definition. For
example, two QLs, $Q(W)$ and $Q(\tau^{-1}W)$ are weakly MLD but not
necessarily SMLD.

A basic lemma as for the MLD relationship is as follows:

\begin{quotation}
\noindent {\bf Lemma}: Let $W$ and $W'$ be the windows of two QLs. Then, a necessary and
sufficient condition for $Q(W')$ to be locally derivable in the strong
sense from $Q(W)$ is that each of the two ends of $W'$ is congruent in
modulo ${\bf Z}[\tau]$ with either end of $W$.
\end{quotation}
This lemma is a direct consequence of the proposition, {\bf P1}, presented in
\S \ref{sec:localenv}. This is also a direct consequence of a
general result presented in ref\verb/./\citen{Ba99}). The following two
propositions are readily derived from this lemma:

\begin{quotation}
\noindent {\bf P2}: Let $W$ and $W'$ be the windows of two QLs. Then, a
necessary and sufficient condition for the two to belong to a common SMLD
class is $|W| \equiv \pm|W'| \bmod {\bf Z}[\tau]$.
\end{quotation}

\begin{quotation}
\noindent {\bf P3}: Let $W$ and $W'$ be the windows of two QLs. Then, a
necessary and sufficient condition for the two to belong to a common MLD
class is $|W| \equiv \pm\tau^n|W'| \bmod {\bf Z}[\tau]$ for some $n \in
{\bf Z}$.
\end{quotation}
It follows from these propositions that all the binary QLs obtained from a
single mother lattice belong to a single MLD class. It is important that a
single MLD class may be divided into several SMLD classes. We may say that
the two QLs in {\bf P2} are properly or improperly SMLD depending on
which case of the two signs $\pm$ is relevant. A similar definition applies
to the MLD relationship and the congruence or scaled-congruence
relationship in modulo ${\bf Z}[\tau]$.

Let $Q(W)$ be a QL and let $W'$ be a subwindow of $W$ associated with a
local environment. Then we can conclude from {\bf P2} that $Q(W')$ is
properly or improperly SMLD if the types of the two ends of $W'$ are
represented as ``lr" or ``rl", respectively. On the other hand, if the types
of the two ends of $W'$ are represented as ``ll" or ``rr", $Q(W)$ is not
locally derivable from $Q(W')$ unless $|W|$ is a $\tau$-integral.

It is appropriate at this point to define exactly self-similarity of QLs; a
self-similarity is a similarity of a geometrical object with a part of
itself. Using the relation $\tau Q(W) \simeq Q(\tau^{-1}W)$, we can show
that two QLs, $Q(W)$ and $Q(W')$, are mutually SLI if $|W'| =
\tau^{-1}|W|$. In particular, if $W'$ is a subset of $W$, $Q(W')$ is a
subset of $Q(W)$. Then, we shall say that $Q(W)$ is weakly self-similar.
Every QL derived from a mother lattice with a hyper-scaling symmetry has
the weak self-similarity and the relevant ratio of self-similarity is equal
to $\tau$. \cite{DuKa85, Ni89a} $Q(W)$ is strongly self-similar if
$Q(W)$ and $Q(W')$ above are SMLD. We will use from now on
``self-similarity" only in this strong meaning. Then, if $Q(W)$ is self-similar,
there exists a natural number $n$ so that $|W'| = \tau^{-n}|W|$ and,
besides, $W$ and $W'$ satisfy the condition in {\bf P2} above. Thus, {\it a
necessary and sufficient condition for $Q(W)$ to be self-similar is
$(\tau^n \pm 1)|W| \in {\bf Z}[\tau]$ for either sign and some natural
number $n$}.
It follows that the window of a QL with a self-similarity must be a
$\tau$-integral or a $\tau$-rational. Therefore, we shall classify a QL
into the type I, II or III according as its window, $|W|$, is a
$\tau$-integral, $\tau$-rational or $\tau$-irrational, respectively; type
III QLs cannot have the self-similarity. Since Nature prefers simpler
structures to complex ones, we shall concentrate our consideration,
hereafter, to QLs of the type I or II. The type II QLs are classified
further into the type II$_{\rm a}$, II$_{\rm b}$ or II$_{\rm c}$ depending
on which type its window as a $\tau$-rational belongs to.

It follows from {\bf P2} and {\bf P3} that all the QLs of the type I form a
single MLD class, which we shall call the fundamental MLD class. The
residue class and the SSRC of a number in ${\bf Q}[\tau]$ in modulo ${\bf
Z}[\tau]$ are essential in a classification of the type II QLs (for SSRC,
see Appendix \ref{Symmetrized scaled residue classes}). It follows from
{\bf P2} that there exists a bijection between the set of all the SMLD
classes of the type II QLs and the set of all the residue classes in ${\bf
Q}[\tau]$. On the other hand, we can conclude from {\bf P3} and the
definition of an SSRC that $\Sigma_{\rm sym.}^+(|W|)\; ( := \Sigma_{\rm
sym.}(|W|) \cap {\bf R}^+ )$ is equal to the set of the windows of all the
QLs in an MLD class to which $Q := Q(W)$ belongs; {\it there exists a
bijection between the set of all the MLD classes and the set of all the
SSRCs.} This is the most important conclusion of the present paper. The
ratio of self-similarity is common among different QLs in the MLD class and
is equal to $\tau^q$ or $\tau^{q'}$ with $q$ or $q'$ being the index or the
half-index of the relevant SSRC.

The set of all the type II$_{\rm a}$ QLs form a single MLD class if $m$ is
odd but it is divided into two MLD classes if $m$ is even. On the contrary,
the set of all the QLs of the type II$_{\rm b}$ is divided into an infinite
numbers of MLD classes. The same is true for the case of type II$_{\rm c}$
QLs.

It is evident that the ratio of self-similarity of the type I QL is equal
to $\tau$. The same is true for the case of type II$_{\rm a}$ QLs
associated with the scaled residue class $C$ with an even $m$. The ratio of
self-similarity of a QL belonging to the scaled residue class
$\Sigma(\frac{1}{2})$ is equal to $\tau^2$ or $\tau^3$ according as $m$ is
even or odd, respectively.

To have a convenient symbol system specifying different MLD classes of type
II QLs is desirable. The relevant mother lattice of an MLD class is
specified by $m^e$ with $m$ and $e$ being the parameters characterizing the
hyper-scaling ratio $\tau$. MLD classes with a common mother lattice form a
family. Another important parameters characterizing an MLD class is the
power $n$ when the ratio of self-similarity of the class is represented as
$\tau^n$; $n$ is equal to the index or the half-index of the relevant SSRC.
We shall call $n$ the scaling index of the MLD class. The three types,
II$_{\rm a}$, II$_{\rm b}$ and II$_{\rm c}$, can be distinguished by the
three signs, $\pm$, $-$ and +, respectively, which appear in the condition
$(\tau^n - \varepsilon)|W| \in {\bf Z}[\tau]$. Two parameters, $n$ and
$\varepsilon$, are specified by $n^\varepsilon$. If there exist more than
one MLD classes for specified $m^e$ and $n^\varepsilon$, we shall
distinguish them by numbering them appropriately. MLD classes with small
scaling indices will be important in a practical application.

Several MLD classes of type II QLs with lower scaling indices are listed in
Table \ref{table1}. There exists no MLD class with scaling index 1 in the
family characterized by $1^-$ because $\tau_{\rm G} + 1 = \tau^2_{\rm G}$
is scaling-equivlent to 1.

\begin{table}
\begin{center}
\begin{tabular}{rccccc}
\hline
{No} & $m^e$ & $n^\varepsilon$ & type $1/\rho$ & type $\omega$ & integral
SIR  \\ \hline
1 & $1^-$ & $2^-$ & $*$ & $\tau^2_{\rm G}/\sqrt{5}\, $ & $2\tau_{\rm
G}/\sqrt{5}\, $  \\
2 & $1^-$ & $3^\pm$ & $\tau^2_{\rm G}/2$ & $\tau^2_{\rm G}/2$ & $*$  \\ \hline
3 & $2^-$ & $1^\pm$ & $1 + 1/\sqrt{2}\, $ & $1 + 1/\sqrt{2}\, $ & $*$  \\
4 & $2^-$ & $2^\pm$ & $\tau/2$ & $*$ & $*$  \\
5 & $2^-$ & $2^-$ & $*$ & $(3 + 2\sqrt{2}\, )/2\sqrt{2}\, $ & $*$  \\ \hline
6 & $4^-$ & $1^\pm$ & $\tau_{\rm G}$ & $\tau_{\rm G}^2$ & $\tau_{\rm G}$  \\
7 & $4^-$ & $1^+$ & $\tau_{\rm G}^2/2$ & $*$ & $*$  \\
8 & $4^-$ & $1^-$ & $*$ & $1 + \tau_{\rm G}/2$ & $*$  \\ \hline
9 & $4^+$ & $1^\pm$ & $(1 + \sqrt{3}\, )/2$ & $(1 + \sqrt{3}\, )/2$ & $*$  \\
10 & $4^+$ & $1^-$ & $*$ & $1 + 2/\sqrt{3}\, $ & $2/\sqrt{3}\, $  \\
11 & $4^+$ & $1^-$ & $*$ & $(5 + 3\sqrt{3}\, )/2\sqrt{3}\, $ & $*$  \\ \hline
12 & $6^+$ & $1^\pm$ & $2 + \sqrt{2}\, $ & $1 + \sqrt{2}\, $ & $\sqrt{2}\, $  \\
13 & $6^+$ & $1^+$ & $(1 + \sqrt{2}\, )/2$ & $*$ & $*$  \\
14 & $6^+$ & $1^-$ & $*$ & $2 + 3/\sqrt{2}\, $ & $1 + 1/\sqrt{2}\, $  \\
15 & $6^+$ & $1^-$ & $*$ & $(3 + \sqrt{2}\, )/2\sqrt{2}\, $ & $*$  \\
16 & $6^+$ & $1^-$ & $*$ & $(7 + 5\sqrt{2}\, )/2\sqrt{2}\, $ & $*$  \\
\hline
\end{tabular}
\end{center}
\caption{A list of MLD classes of type II QLs with low scaling indices.
The last three columns show the windows of inflation-symmetric QLs of
specified types as explained in a later section.}
\label{table1}
\end{table}

A remained task is to investigate mutual relations among different QLs in a
single MLD class. This will be carried out in the subsequent several
sections.

\section{Subquasilattice and Decoration Rule}\label{sec:DR}

The LR-division of the window $W$ of a ternary QL defines six boundary
points of the five subwindows. Let $W'$ be a subwindow whose two ends
coincide with two of the six boundary points. Then, $Q' := Q(W')$ is
derived from $Q$ by deleting every point which is in one of the particular
LR-environments. If, for example, $W' := W - [1\bullet\rho]$ for the case A
in eq\verb/./(\ref{eqn:mld17}), $Q(W')$ is derived from $Q$ by decimating
all the lattice points in the local environment $[1\bullet\rho]$. On this
decimation, every interval of type $\rho$ merges with an interval of type 1
being located on the left, yielding a new interval of the length $1 + \rho$
which is equal to the length of $\omega$. Hence $Q'$ is a binary QL
composed of the two types of intervals $1$ and $\omega$; $|W'| = \sigma$.
Conversely, $Q$ is retrieved from $Q'$ by decorating a part of intervals of
the type $\omega$ as $1\bullet\rho$ but others are retained intact.
Unfortunately, the rule by which the intervals to be decorated are chosen
are not local for generic $\delta$. That is, $Q'$ is locally derivable from
$Q$ but the converse is not necessarily true. This is consistent with the
fact that the subwindow $W'$ is of the type ``ll".

We next choose the following subwindow for the case $|W| \in A$ in
eq\verb/./(\ref{eqn:mld17}):
\begin{equation}
W' = [1\bullet1] \vee [\omega\bullet 1] \vee [\rho\bullet 1] = [*\bullet1]
\label{eqn:mld19}
\end{equation}
with $|W'| = |W| - 1$. This case satisfies {\bf P2}. We shall
investigate more closely the MLD relationship between the relevant two QLs.
By definition, $Q'$ is derived from $Q$ by decimating all the lattice
points in the LR-environment $1\bullet\rho$ or $1\bullet\omega$ in $Q$ or,
equivalently, by retaining every lattice point which is in the environment $*\bullet
1$. On this decimation, new intervals, $\rho' := 1\rho$ and $\omega' :=
1\omega$, emerge and all the intervals of the types $\rho$ and $\omega$ in
$Q$ disappear. However, a part of the intervals of the type $1$ survive,
and $Q'$ is a ternary QL composed of the three types of intervals $1' :=
1$, $\rho'$ and $\omega'$. Conversely, $Q$ is retrieved from $Q'$ by
decorating every interval of the type $\rho'$ as $1\bullet\rho$ and that of
$\omega'$ as $1\bullet\omega$. Thus, two QLs, $Q$ and $Q'$, are MLD and,
besides, $Q'$ is a subset of $Q$. Then, we shall call $Q'$ a {\it
subquasilattice} of $Q$ and denote it as $Q \succ Q'$. Note that the
frequency vector of $Q'$ is given by $w' = {}^{\rm t}(\sigma - \delta - 1,
\;1 - \delta, \;\delta)$ (cf. eq\verb/./(\ref{eqn:mld16b})) because the frequency of ``1" decreases by 1 on
the decimation.

As noted above the three types of intervals in $Q'$ are related to those in
$Q$ by
\begin{equation}
{\rm A:} \quad 1' = 1, \;\rho' = 1\rho, \;\omega' = \;1\omega.
\label{eqn:mld19a}
\end{equation}
A set of such relations defined between a QL and its subquasilattice is
called a decoration rule (DR). \cite{FootNote8}
The interval (resp\verb/./ frequency) vector of $Q'$ is related to that of
$Q$ by $u' = uL_{\rm A}$ (resp\verb/./ $w' = L^{-1}_{\rm A}w$), where
$L_{\rm A}$ is the decoration matrix defined by
\begin{equation}
 L_{\rm A} = \left( \begin{array}{ccc}
        1 & 1 & 1  \\
        0 & 1 & 0  \\
        0 & 0 & 1
\end{array} \right). \label{eqn:mld19b}
\end{equation}
Note that $L_{\rm A} \in GL^+(3, \;{\bf Z})$.

A general expression for the DR combining a QL, $Q$, and one of its
subquasilattice, $Q'$, is written as $1' = S_1$, $\rho' = S_\rho$ and
$\omega' = S_\omega$, where $S_x$ with $x = 1, \;\rho$ and $\omega$ are
strings of the three kinds of intervals $1, \;\rho$ and $\omega$. The
three strings form a ``vector" ${\bf S} := (S_1, \;S_\rho, \;S_\omega)$,
which we shall call the decoration vector, and a DR is represented as $u' =
{\bf S}$.

\begin{figure}
\begin{center}\epsfile{file=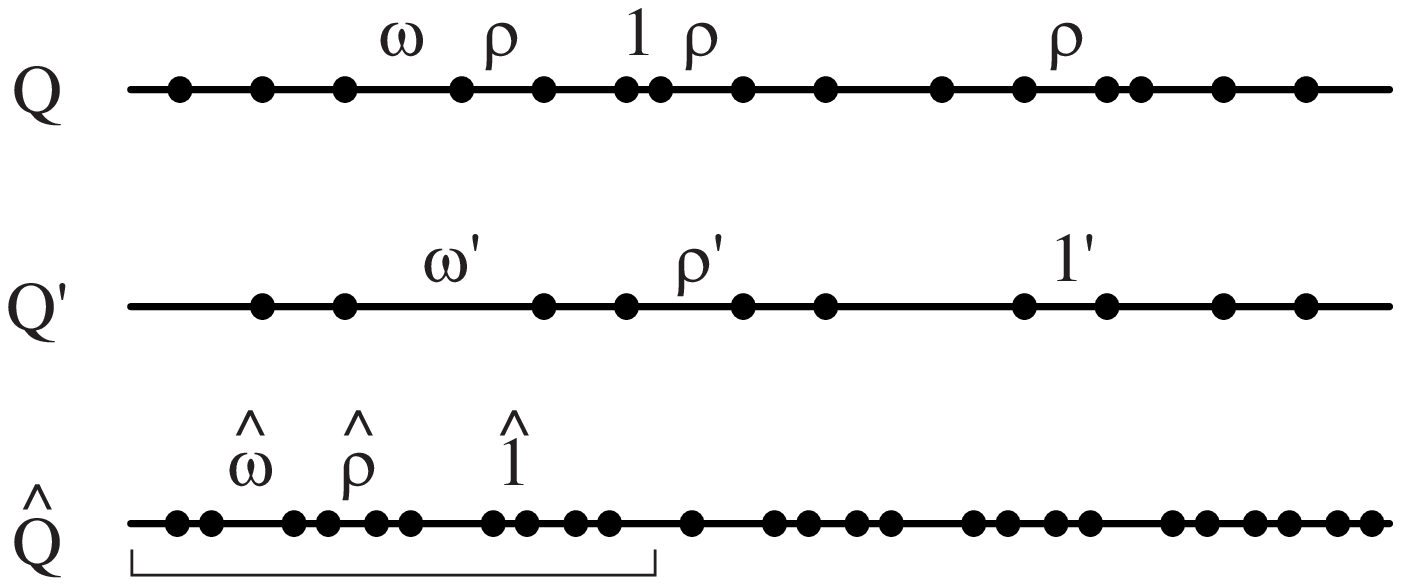,height=4.8cm}\end{center}
\caption{A QL, $Q$, and its subquasilattice, $Q'$, which turns to the third
QL, $\hat{Q}$, by scaling down with the ratio $\tau^{-1}$, where
$\tau=1+\sqrt{2}\, $. The relevant window sizes are:
$|W|=\frac{1}{2}(1+\sqrt{2}\, )$, $|W'|=\frac{1}{2}(3-\sqrt{2}\, )$ and
$|\hat{W}|=\frac{1}{2}+\sqrt{2}\, $. Note that $\hat{1}=1$,
$\hat{\rho}=\sqrt{2}\, $, $\rho=\hat{\omega}=1+\sqrt{2}\, $ and
$\omega=2+\sqrt{2}\, $. A mark is placed in $\hat{Q}$ where a shrinked
version of $Q'$ can be seen.}
\label{fig:Decoration}
\end{figure}

A subquasilattice of $Q$ is obtained also for the case of $|W| \in B$ and
$|W| \in C$ if the subwindow $W'$ is chosen as follows:
\begin{equation}
     \begin{array}{ll}
       W' = [\omega\bullet1] \vee [\rho\bullet 1]  =  [*\bullet 1], & |W|
\in B, \\
       W' = [1\bullet\rho] \vee [\omega\bullet\rho] \vee [\rho\bullet\rho]
 =  [*\bullet\rho], & |W| \in C.
     \end{array} \label{eqn:mld20}
\end{equation}
Note that $|W'| = |W| - 1$ for the case $|W| \in B$ but $|W'| = |W| -
\sigma$ with $\sigma = 1/\tau$ for the case $|W| \in C$. The decoration
vectors are given by
\begin{equation}
     \begin{array}{cl}
       {\rm B:} & (1\rho, \;1\omega^r, \;1\omega^{r+1}), \\
       {\rm C:} & (\rho, \;\rho 1, \;\rho\omega);
     \end{array} \label{eqn:mld20a}
\end{equation}
the repetition number $r$ of $Q$ has appeared in the case B because the
lattice points in the environment $[\omega\bullet\omega]$ are decimated. A
QL and its subquasilattice associated with the DR C are shown in
Fig\verb/./\ \ref{fig:Decoration}. Since $\rho = \tau - 1$ for the case B
and $\rho = \tau$ for the case C, we obtain $1' = \tau$ for these two
cases. This is the reason why the interval $1'$ for the case B or C has
chosen as indicated above. Remember that $\omega = \tau$ for the case B.
The relevant decoration matrices of the cases B and C are given by the
first two of the following three:
\begin{eqnarray}
&& L_{\rm B} =
   \left( \begin{array}{ccc}1 & 1 & 1  \\ 1 & 0 & 0  \\ 0 & r & r+1  \\
          \end{array}\right), \quad
          L_{\rm C} =
   \left( \begin{array}{ccc}0 & 1 & 0  \\ 1 & 1 & 1  \\ 0 & 0 & 1  \\
          \end{array}\right), \quad
          L_{\rm D} =
   \left( \begin{array}{ccc}1 & 0 & 0  \\ 1 & 1 & 1  \\ 0 & r & r+1  \\
          \end{array}\right).\label{eqn:mld20b}
\end{eqnarray}

If the subinterval $D$ is relevant, the subwindow $W'$ is chosen as
\begin{equation}
       \qquad  W' =  [1\bullet\rho] \vee [\omega\bullet\rho] =
[*\bullet\rho], \quad |W| \in D \label{eqn:mld20c}
\end{equation}
with $|W'| = |W| - \sigma$, and the relevant decoration vector is given by
\begin{equation}
\quad {\rm D:} \quad (\rho 1, \;\rho\omega^r, \;\rho\omega^{r+1}),
\label{eqn:mld20d}
\end{equation}
while the decoration matrix is given by the last of the three above. In
fact, only the case $r = 1$ appears for $e = -1$ because the irrational
associated with the QLs for which the case D  is relevant is the golden
ratio; a general case is relevant to the case $e = 1$ to be discussed
shortly.

Two interval vectors $u$ and $u'$ which are combined by a decoration matrix
$L$ as $u' = uL$ are orthogonal to the column vector ${}^{\rm t}(1, \;1,
\;-1)$ on account of the addition rule. Therefore the column vector must be
an eigenvector of $L$. The corresponding eigenvalue must be equal to 1 or
$-$1 because $L$ is unimodular. We shall call the eigenvalue the signature
of $L$ or of the relevant DR. The signatures of the DRs of the four cases,
A, B, C and D, above are plus. It can be shown that the signature
of a DR is plus (resp\verb/./ minus) if the QL and its subquasilattice are
properly (resp\verb/./ improperly) MLD. \cite{FootNote5}

The new window $|W'|$ belongs to $F$ for the case $|W| \in A$. This is not
the case for $|W| \in B$ and $|W| \in C$ because $|W'| < 1$ but $|{\hat W}|
:= \tau |W'|$ belongs to $F$. More precisely, $|{\hat W}| = \tau(|W| - 1)$
or $|{\hat W}| = \tau(|W| - \sigma)$ for the case B or C, respectively.
Using the relation $1' = \tau$, we can conclude that ${\hat Q} := Q({\hat
W})$ $(= \tau^{-1} Q')$ is a ternary QL with three types of new intervals
1, ${\hat \rho}$ and ${\hat \omega}$ with ${\hat \rho} = \tau^{-1}\rho'$
and ${\hat \omega} = 1 + {\hat \rho}$, as shown in Fig\verb/./\
\ref{fig:Decoration}. Since $\omega = \tau$ and $\rho' = 1\omega^r$ for the
case B, we obtain the equation ${\hat \rho} = \tau^{-1} + r$, which
determines the repetition number as $r = m - {\hat k}$ with ${\hat k}$
being the class number of ${\hat Q}$: $\hat{\rho} = \tau - \hat{k}$. The
repetition number for the case D will be determined later on.

We can choose a different subquasilattice of $Q$ from above for the case of
$|W| \in B$ if the subwindow $|W'|$ is chosen as
\begin{equation}
       W' = [1\bullet\omega] \vee [\omega\bullet\omega] = [*\bullet\omega],
\quad |W| \in B,
     \label{eqn:nega1}
\end{equation}
so that $|W'| = 1 + \sigma - |W|$ with $\sigma = 1 + 1/\tau$. The
decoration vector assumes
\begin{equation}
      {\rm B':} \; (\omega, \;\omega 1(\rho 1)^r, \;\omega 1(\rho1)^{r+1})
     \label{eqn:nega2}
\end{equation}
with $\rho = \tau - 1$ and $r = m - 1 - {\hat k}$, while the the decoration
matrix is given by the left one of the following two:
\begin{eqnarray}
&& L_{\rm B}' =
   \left( \begin{array}{ccc}0 & r + 1 & r + 2  \\ 0 &  r & r + 1  \\ 1 & 1
& 1  \\
          \end{array}\right), \quad
          L_{{\rm D}'} =
   \left( \begin{array}{ccc}0 &  r & r + 1  \\ 0 & r + 1 & r + 2  \\ 1 & 1
& 1  \\
          \end{array}\right).\label{eqn:nega3}
\end{eqnarray}
The QL and its subquasilattice are improperly MLD for this case in contrast to
other cases above, and the signature of the DR is minus. The same is true for
the case ${\rm D}'$ below.

A binary QL is considered to be a limiting ternary QL where the frequency
of the longest interval, $\omega$, vanishes. Consequently, most result
derived for the case of a ternary QL is basically applicable to the case of
a binary QL. More precisely, only the cases A and C appear for this case,
and the relevant decoration matrices are related to the binary counterparts
given in \S \ref{sec:binary} as $L_{\rm A} = \left({L \atop 0}\; {*
\atop 1}\right)$ and $L_{\rm C} = \left({L' \atop 0}\; {* \atop 1}\right)$.

We shall proceed to the case of $e = 1$. It can be shown that the DRs for
the cases A and D are the same as the corresponding results for the case of
$e = -1$ but a slight modification is necessary for the case B, C or ${\rm
B}'$ on account of the fact that $\rho$ can take a value smaller than 1.
The resulting decoration vectors for the case B, C and ${\rm B}'$ assume
\begin{equation}
     \begin{array}{cl}
       {\rm B:} & (1\omega, \;1\rho, \;1\omega\omega), \\
       {\rm C:} & (\rho 1, \;\rho, \;\rho\omega), \\
       {\rm B':} & (\omega 1, \;\omega, \;\omega 1\rho 1),
     \end{array} \label{eqn:mld20e}
\end{equation}
respectively, because the length of $1'$ must be equal to $\tau$. Consequently,
the modified DR for the case C becomes identical to the DR for the case D
but with $r = 0$. The relevant decoration matrix is written as $L_{\rm C} =
\left({L'' \atop 0}\; {* \atop 1}\right)$ with $L''$ being the 2D matrix
given in \S \ref{sec:infsym}. It can be shown that the repetition
number in the case D is given by $r=m - 1 - \hat{k}$ with $\hat{k}$ being
the class number of $\hat{Q}$. Remember that the DRs for the cases C and D
can be treated in a unified way for the present case of $e = 1$.

A similar DR to ${\rm B}'$ is obtained for the case of $|W| \in D$ if the
subwindow $|W'|$ is chosen as
\begin{equation}
       W' = [\omega\bullet\omega] \vee [\rho\bullet\omega] =
[*\bullet\omega], \quad |W| \in D = \;]1, \;2\sigma] \label{eqn:nega4}
\end{equation}
with $\sigma = 1 - 1/\tau$, so that $|W'| = 1 + \sigma - |W|$. The
decoration vector is given by
\begin{equation}
      {\rm D':} \; (\omega, \;\omega \rho(1\rho)^r, \;\omega
\rho(1\rho)^{r+1})
     \label{eqn:nega5}
\end{equation}
with $\rho = \tau - 1$ and $r = m - 2 - {\hat k}$, while the the decoration
matrix is given by the right one of eq\verb/./(\ref{eqn:nega3}). Remember
that a similar DR is possible for the case $e = -1$ and $\tau = \tau_{\rm
G}$.

\section{Inflation-Symmetric QLs}\label{sec:infs}

\subsection{Saw map}\label{sec:SM}

It is sometimes neccessary to bring a window, $W$, with a generic size into the
fundamental interval $F$ by scaling it as $\tau^nW$ with ${}^\exists n \in
{\bf Z}$. This procedure defines a map, ${\rm Sc}(*)$, from ${\bf R}^+$
onto $F$. The map is a discontinuous but piecewise linear function with the
scaling symmetry: ${\rm Sc}(\tau x) = {\rm Sc}(x)$.

\begin{figure}
\begin{center}\epsfile{file=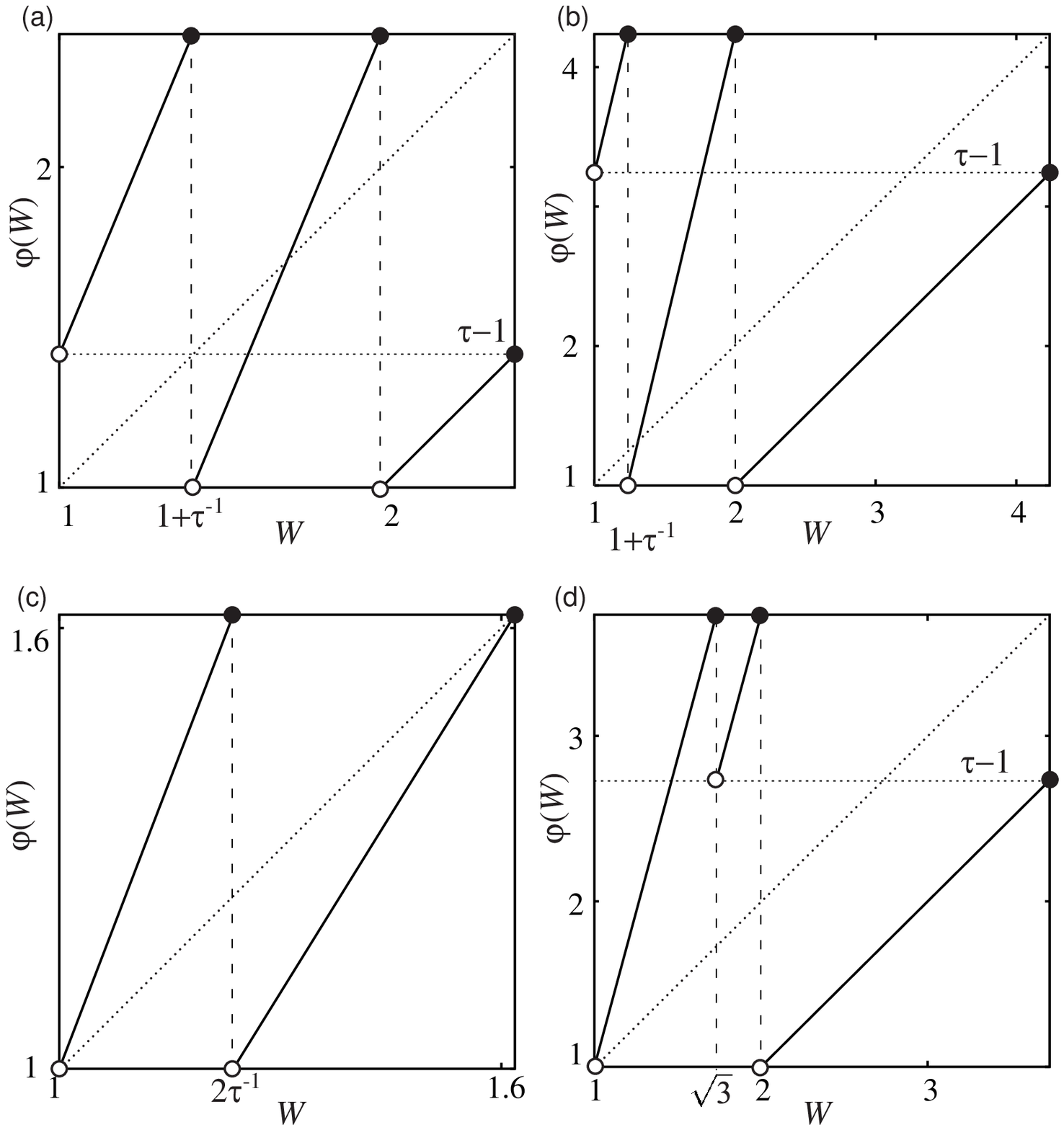,height=11cm}\end{center}
\caption{Saw maps, $\varphi(W)$, are shown in solid lines for four values
of $\tau$: (a) $1+\sqrt{2}\, $, (b) $2+\sqrt{5}\, $, (c) $\tau_G$ and (d)
$2+\sqrt{3}\, $. The scales of the boxes are so chosen that the fundamental
intervals, $F$, are normalized. An SM maps $F$ onto itself. The inverse map
is two-valued. The fixed points are given by the condition $\varphi(W)=W$
(dotted line). A cycle of an SM yields a set of ISQLs which are MLD.}
\label{fig:SawMapping}
\end{figure}

In what follows we shall usually identify window $W$ with its size $|W|$.
We have defined in the preceding section a transformation of window $W$
into another one ${\hat W}$ when $W$ belongs to the subinterval B or C. If
we define ${\hat W} := W' = W - 1$ for the case of $W \in {\rm A}$, we may
write ${\hat W} = \varphi(W)$ for $W \in F$ with $\varphi(W)$ being a map
from $F$ onto $F$:
\begin{equation}
 \varphi(W) = \left\{
     \begin{array}{ll}
        {\rm Sc}(W - 1), & W \in \Gamma^{\rm c}, \\
        {\rm Sc}(W - \eta), & W \in \Gamma,
     \end{array} \right. \label{eqn:mld20f}
\end{equation}
with $\eta := 1/\tau$. To be more specific, we obtain
\begin{equation}
 \varphi(W) = \left\{
     \begin{array}{ll}
        W - 1, & W \in {\rm A} = \;] 2, \;\tau], \\
        \tau(W - 1), & W \in {\rm B} = \;] 1 + \eta, \;2], \\
        \tau W - 1, & W \in {\rm C} = \;] 1, \;1 + \eta].
     \end{array} \right. \label{eqn:mld21}
\end{equation}
The map is discontinuous, and composed of three linear pieces as shown in
Fig\verb/./\ \ref{fig:SawMapping}. Every linear piece of the map has a
positive slope whose value is 1 or $\tau$. We shall call the map a saw map
(SM).

The rescaling does not participate in the A region of the SM, and $W$
decreases there, while that does in the B and C regions. Let $n$ be the
number of mapping procedures which are needed in order to reach the
interval B or C when we start from A. Then, $W$ is changed into $W
- n$ by the $n$ step procedure, and the corresponding decoration vector
assumes $(1, \;1^n\rho, \;1^n\omega)$. The $n$ step procedure may be called
the cascade process. The step number $n$ is equal to $\lceil W\rceil - 1$.
The window escapes the interval A by a single cascade process, which
bypasses $n -1$ steps of mapping procedures.

If $Q \succ Q'$ and $Q' \succ Q''$, then $Q \succ Q''$. We shall consider, as
an example, the case $\tau = 1 + \sqrt{2}\, $ ($m = 2$). If $W \in {\rm A}$,
then $Q(W) \succ Q(W') \succ Q(W'')$ for $W' := W - 1$ and $W'' := W' -
\tau^{-1}$. The DR combining the three types of intervals of $Q''$ with
those of $Q$ is a composition of the DR between $Q$ and $Q'$ and that of
$Q'$ and $Q''$, so that we obtain $1'' = 1\rho$, $\rho'' = 1\rho 1$ and
$\omega'' = 1\rho 1\omega$. The corresponding interval vectors are mutually
related by $u' = uL_{\rm A}$, $u'' = u'L_{\rm C}$ and $u'' = uL$ with
$L_{\rm A}$, $L_{\rm C}$ and $L$ being the relevant decoration matrices.
Therefore, we obtain a composition rule $L = L_{\rm A}L_{\rm C}$. A
composition of DRs is possible also for the case of more than three
generations of subquasilattices. A composite DR is composed of the DRs
given in \S \ref{sec:DR}, and we may call these simple DRs. The
length of the first component, $S_1$, of a decoration vector ${\bf S}$ is
a power of $\tau$. We shall call the power the scaling order of the
relevant DR. The scaling order of a simple DR is equal to 0 or 1. The
scaling order of a composite DR of two DRs is the sum of those of the
latters, while the signature of the composite DR is equal to the products
of the signatures of the latters.

If a QL, $Q$, and its subquasilattice $Q'$ are mutually SLI (scaling-LI),
$Q$ has an inflation-symmetry. The DR between $Q$ and $Q'$ in this case is
nothing but the inflation rule (IR), so that the inflation matrix is
considered to be a special decoration matrix. The scaling order is defined
also for an IR; the ratio of inflation-symmetry of the relevant QL is equal
to $\tau^n$ with $n$ being the scaling order, which satisfies $n \ge 1$

We have been assumed implicitly that $e = -1$ and $m \ge 2$ but
eq\verb/./(\ref{eqn:mld20f}) remains valid for other cases as well provided
that we take $\eta = 1 - 1/\tau$ if $e = 1$. Note, however, that the first
line of eq\verb/./(\ref{eqn:mld20f}) is unnecessary if $m = 1$ because
$\Gamma^{\rm c}$ is empty then. It follows that the explicit form of $\varphi(W)$
for $e = -1$ and $m = 1$ (i. e., $\tau = \tau_{\rm G}$) assumes:
\begin{equation}
 \varphi(W) = \left\{
     \begin{array}{ll}
        \tau W - 1, & W \in {\rm C} = \;] 2\tau^{-1}, \;\tau], \\
        \tau^2 W - \tau, & W \in {\rm D} = \;] 1, \;2\tau^{-1}],
     \end{array} \right. \label{eqn:mld22}
\end{equation}
while the third line of eq\verb/./(\ref{eqn:mld21}) should be replaced for
$e = 1$ by
\begin{equation}
 \varphi(W) = \tau(W - 1) + 1, \quad W \in {\rm C}.
 \label{eqn:mld22b}
\end{equation}
The SM for the case of $\tau = \tau_{\rm G}$ or $\tau = 2 + \sqrt{3}\, $ is
shown in Fig\verb/./\ \ref{fig:SawMapping}(c) or (d), respectively.

We have assumed tacitly that the DR for the case B is given by the first
row of eq\verb/./(\ref{eqn:mld20a}). If we adopt
eq\verb/./(\ref{eqn:nega2}), the SM must be modified in the interval B as
\begin{equation}
 \varphi(W) = \tau(2 - W) + 1, \quad W \in {\rm B}.
 \label{eqn:mld22c}
\end{equation}
Similarly, if we adopt eq\verb/./(\ref{eqn:nega5}) for the case D, the
relevant SM for the case $e = 1$ assumes
\begin{equation}
 \varphi(W) = \tau(2 - W) - 1, \quad W \in {\rm D}.
 \label{eqn:nega6}
\end{equation}
but that for the case $e = -1$ and $\tau = \tau_{\rm G}$ assumes
\begin{equation}
 \varphi(W) = \tau^3 - \tau^2 W, \quad W \in {\rm D}.
 \label{eqn:nega7}
\end{equation}
If $e = -1$, the modified SM is continuous and its form resembles the
letter ``{\it N}\,", so that we shall call it NSM not only for $e = -1$ but also
for $e = 1$. The NSM has basically similar properties to those of the SM.
An important property of the NSM is that it combines two scaled residue
classes forming an SSRC of type II$_{\rm c}$.

\subsection{Inflation symmetry}\label{sec:inf}

If the SM eq\verb/./(\ref{eqn:mld21}) is repeated indefinitely, we obtain
an infinite series of windows: $W^{(0)} \rightarrow W^{(1)} \rightarrow
W^{(2)} \rightarrow \cdots$, where $W^{(0)} := W$ and $W^{(n)} =
\varphi(W^{(n - 1)})$ ($n \ge 1$). This series of windows define a
semi-infinite chain of subquasilattices, $Q_0 \succ Q_1 \succ Q_2 \succ
{}\cdots$ with $Q_0 := Q$, provided that scales of these QLs are ignored.
It is obvious that the series of subquasilattices generated by the SM are
properly MLD. By the properties of the SM, the SM never has any stable
fixed point, and the series is chaotic if $W$ is a $\tau$-irrational.

We can prove that the series of windows generated by the SM from an initial
window belonging to ${\bf Q}[\tau]$ has the following properties:
\begin{enumerate}
\def\labelenumi{\theenumi}
\def\labelenumi{\theenumi)}
\def\theenumi{\roman{enumi}}
\item The series of windows always falls eventually into a cycle.

\item The cycle is determined by the scaled residue class to which the
initial window belongs.

\end{enumerate}
Note that cycles which are equivalent with respect to a cyclic permutation
are not distinguished in ii) above.
Exactly speaking, the item ii) is justified only for the case of MLD
classes of the type II$_{\rm a}$ or II$_{\rm b}$ because the SM does not
combine two scaled residue classes in an SSRC of type II$_{\rm c}$. This
difficulty is resolved, however, by the use of the NSM.
Proofs of the two properties above are
not presented here but will be presented elsewhere because a
considerable space is required.

If the cascade process is used in the A region of the SM, the
above-mentioned series of windows is modified because there exist bypassed
windows. The cascade process shortens the period of a cycle but it does not
essentially change the cycle.

There can be two cases; the series is purely cyclic or mixed-cyclic, in
which the series falls into a cycle after finite terms. For example, a pure
3-cycle and a pure 4-cycle are obtained for the windows listed in the
fourth row and the tenth one of Table \ref{table1}:
\begin{equation}
\frac{1}{2}(1 + \sqrt{2}\, ) \in C \rightarrow \frac{1}{2} + \sqrt{2}\, \in
B \rightarrow \frac{3}{2} + \frac{1}{2}\sqrt{2}\, \in A \rightarrow \;
\cdots, \label{eqn:mld21b}
\end{equation}
\begin{equation}
\frac{2}{\sqrt{3}\, } \in D \rightarrow 1 + \frac{1}{\sqrt{3}\, } \in C
\rightarrow 2 +
\frac{2}{\sqrt{3}\, } \in A \rightarrow 1+ \frac{2}{\sqrt{3}\, } \in A
\rightarrow \; \cdots.
\label{eqn:mld21c}
\end{equation}
We shall call the length of a cycle the period. A window is called cyclic
if it yields a purely cyclic series of windows.

If composition of an SM is repeated $n$ times, we obtain an $n$-fold SM
$\varphi^{n}$. A fixed point of $\varphi^{n}$, which satisfies $W =
\varphi^{n}(W)$, is called a proper fixed point if it is not a fixed point
with a lower period, which must be a divisor of $n$. A proper fixed point
of $\varphi^{n}$ is cyclic, and its period is equal to $n$. The converse of
this statement is true as well. The $n$-fold SM is also an SM, and the
number of its peaks increases exponentially as a function of $n$.
Therefore, the number of the fixed points will increase rapidly as well.
The points of discontinuity of $\varphi$ as well as those of $\varphi^{n}$
belong to ${\bf Z}[\tau]$.

In a purely cyclic case, all the QLs in the relevant chain of
subquasilattices have inflation-symmetries. The number of different QLs in
the series is equal to the period of the series (or chain). We have already
an algorithm of deriving the IR when a window in a cycle is given. From now
on we shall use the abbreviation ``ISQL" for ``inflation-symmetric QL". The
ISQLs in one cycle are cyclically related to each other by DRs.

We shall call the decoration vector an inflation vector if it represents an
IR. For example, the inflation vector representing the IR for the QL whose
window is the first of the 3-cycle (\ref{eqn:mld21b}) is given by
\begin{equation}
(\rho^2 1, \;\rho^2 1\rho^2\omega, \;\rho^2 1\rho^2\omega\rho\omega)
\label{eqn:mld21d}
\end{equation}
with $\rho = 1 + \sqrt{2}\, $. The ratio of inflation-symmetry of this QL is
equal to $\tau^2$, which is the length of the interval $1' \;(= \rho^2 1)$,
while the frequency vector is proportional to $(1, \; 1 + 2\sqrt{2}\, ,
\;1)$.

In a mixed-cyclic case, one cycle of the cyclic part of the series (or
chain) define several ISQLs, which are subquasilattices of the initial QL.
Therefore, the initial QL is a decoration of an ISQL in the chain.

We can derive several conclusions from the two results above:
\begin{enumerate}
\def\labelenumi{\theenumi}
\def\labelenumi{\theenumi)}
\def\theenumi{\roman{enumi}}
\item Every MLD class of QLs includes a finite number of ISQLs.

\item We can choose an ISQL as the representative of an MLD class; other
QLs in the class are given as decorations of the representative.
\end{enumerate}

An SM, $\varphi$, is defined for each quadratic irrational $\tau$
characterizing the mother lattice of 1DQLs. Since it is not monotonic, the
inverse SM, $\varphi^{-1}$, is two-valued, so that $\varphi^{-n} :=
(\varphi^{-1})^n$ with $n$ being a natural number is $2^n$-valued. Hence,
if the initial window is fixed, the inverse SM generates a tree. Since
every window belonging to a single scaled residue class generates a series
which falls in a common cycle for the case of MLD classes of the type
II$_{\rm a}$ or II$_{\rm b}$, all the windows of the class form an inverted
tree, and the same is true for QLs in a single MLD class. In other words, the
inverted tree is like a river system in which all the branches eventually
merge to form the main stream. Naturally, two trees must be
considered for the case of the type II$_{\rm c}$.

Since different ISQLs belonging to a single MLD class of QLs have different
IRs, the IR is not a good index to discriminate different MLD classes.

An important property of the NSM is that it may yield an ISQL with the
minus signature. A QL of the type II$_{\rm a}$ can have two types of the
IRs with different signatures, while the IR of an ISQL of the type II$_{\rm
c}$ must have a plus signature. On the other hand, the IR of an ISQL of the
type II$_{\rm b}$ has a plus or a minus signature according as the scale of
the inflation symmetry is $\tau^q$ or $\tau^{q'}$, respectively, where $q$
or $q'$ is the index or the half-index of the relevant scaled residue
class.

The arguments in the present section are concentrated mainly to the SM but
similar arguments are possible for the NSM.

\subsection{Inflation matrix}\label{infmtrx}

Let $M \in GL^+(3, \;{\bf Z})$ be an inflation matrix of a QL with the
window $W \in {\bf Q}^+[\tau]$. Then it has three eigenvalues $\tau^n$,
${\bar \tau}^n$ and $\varepsilon$, where $n$ stands for the scaling order
and $\varepsilon \;(= \pm1)$ the signature. The first of the three is the
Frobenius eigenvalue, and the left Frobenius eigenvector is given by the
interval vector $u$; $\tau^n u = u M$. The frequency vector $w$ of the
relevant QL is proportional to the right Frobenius eigenvector; it is
written with $W$ as
\begin{equation}
w = {}^{\rm t}(W - 1, \;W - \sigma, \;1 + \sigma - W), \label{eqn:infmtr1}
\end{equation}
whose components are $\tau$-rationals.

The right eigenvector corresponding to the eigenvalue $\varepsilon$ is
given by ${}^{\rm t}(1, \;1, \;-1)$, which must be orthogonal to the
interval vector $u$; this leads to the addition rule for a ternary QL. We
can choose a row vector with integer components as a left eigenvector of
$M$ for the eigenvalue $\varepsilon$. The orthogonality of this vector to
$w$ yields a linear relation with integral coefficients among the
components of $w$; presence of the linear relation is a direct consequence
of that the three components of $w$ belong to ${\bf Q}[\tau]$. For example,
the left eigenvector assumes $(1, \;0, \;-1)$ if $W = 1
+ \sigma/2$ with $1 < \sigma < 2$ because the frequencies of the intervals
1 and $\omega$ are equal then.

We can show readily that ${\rm det}\, M = \varepsilon e^n$ and ${\rm Tr}\, M
= N + \varepsilon$ with $N = \tau^n + {\bar \tau}^n$ being a natural
number. The number $N$ is odd only when $m$ is odd and, besides, $n$ is not
a multiple of 3. The characteristic polynomial $P(x) := {\rm det} (xI - M)$
is factorized as
\begin{equation}
P(x) = (x - \varepsilon)(x^2 - Nx + e^n) \label{eqn:infmtr1b}
\end{equation}
because $\tau^n$ satisfies the quadratic equation, $x^2 - Nx + e^n = 0$.
The Cayley-Hamilton theorem verifies $P(M) = 0$.

If each matrix element of $M$ is replaced by 0 or 1 according as its parity
is even or odd, respectively, we obtain $M \bmod 2$, i.e., $M$ in modulo 2.
It has the column vector ${}^{\rm t}(1, \;1, \;1)$ as its eigenvector
because ${}^{\rm t}(1, \;1, \;1) \equiv {}^{\rm t}(1, \;1, \;-1) \bmod 2$.
Therefore, the number of 1 in the three components of each row vector of $M
\bmod 2$ must be odd. Moreover, the three row vectors of $M$ must be all
different because ${\rm det}M \equiv 1 \bmod 2$. Hence, there can be two
cases: i) $M \bmod 2$ is equal to a permutation matrix or ii) it is derived
from a permutation matrix by replacing a row by the row vector $(1, \;1,
\;1)$. Using the Cayley-Hamilton theorem, we can show that $M^3 \equiv I
\bmod 2$ or $M^4 \equiv I \bmod 2$ according as $N$ is odd or even,
respectively. Therefore, $M \bmod 2$ turns to a permutation matrix if it is
repeatedly multiplied by itself at most three times.

The window satisfies $(\tau^n \pm 1)W \in {\bf Z} [\tau]$ for either sign.
Using this and the fact that $w$ and ${}^{\rm t}(1, \;1, \;-1)$ are right
eigenvectors of $M$ corresponding to the eigenvalues $\tau^n$ and
$\varepsilon$, respectively, we can conclude that $(\tau^n - \varepsilon)W$
is equal to a $\tau$-integral. As a consequence, it follows from
eq\verb/./(\ref{eqn:infmtr1}) that all the components of the scaled
frequency vector $(\tau^n - \varepsilon)w$ are $\tau$-integrals. Thus,
$\varepsilon$ can be negative only when $W$ is a $\tau$-rational of type a
or b.

Using recursively the Cayley-Hamilton theorem, $P(M) = 0$, we obtain
\begin{equation}
M^i = P_iM^2 + Q_iM + R_iI, \label{eqn:infmtr2}
\end{equation}
where $P_i$, $Q_i$ and $R_i$ are integers. Multiplying $M$ on both sides
of this equality and using $P(M) = 0$ again, we can relate $Q_i$'s and
$R_i$'s with $P_i$'s by the equations:
\begin{equation}
Q_i = P_{i+1} - (N + \varepsilon)P_i, \quad R_i = \varepsilon e^nP_{i-1}.
\label{eqn:infmtr3}
\end{equation}
The series $P_i$ ($i \in {\bf Z}$) satisfies a three-term recursion
relation associated with the polynomial $P(x)$ as given by
eq\verb/./(\ref{eqn:infmtr1b}). It is expressed as a linear combination of
three fundamental solutions, one of which is assumed to be $\varepsilon^i$
($i \in {\bf Z}$), while the remaining two are assumed to be two
fundamental solutions of the two-term recursion relation $X_{i+1} = NX_i -
e^nX_{i-1}$. Let $G_i$ be a solution of this recursion relation with the
initial conditions, $G_0 = 0$ and $G_1 = 1$. Then, $G_i$ and its shifted
form, $G_{i-1}$, yield two fundamental solutions. The three coefficients in
the expression for $P_i$ are determined by the initial conditions, $P_0 =
P_1 = 0$ and $P_2 = 1$, obtaining
\begin{equation}
P_i = \frac{G_i - \varepsilon e^nG_{i-1} - \varepsilon^{i-1}}{N -
\varepsilon(1 + e^n)}, \label{eqn:infmtr4}
\end{equation}
which is an integer contrary to its appearence. $Q_i$ and $R_i$ are also
expressed as linear combinations of the three fundamental solutions, and so
is every matirx element of $M^i$. Note that $G_i = F_{ni}/F_n$ and $N =
F_{n+1} - eF_{n-1}$, where $F_i$'s are the generalized Fibonacci numbers;
$F_i$'s are identical to $G_i$'s defined for the case of $n = 1$.

If the inflation rule of a type II ISQL is applied $i$ times to the
interval vector $u = (1, \;\rho, \;\omega)$, we obtain the interval vector
in the $i$-th generation: $u^{(i)} = (1^{(i)}, \;\rho^{(i)},
\;\omega^{(i)})$, which is represented in two ways as $u^{(i)} = uM^i =
\tau^nu$. Let $X$ be one of the three intervals in the $i$-th generation.
Then a periodic lattice whose one period is equal to $X$ is a periodic
approximant to the ISQL. The numbers of the three types of intervals
included in $X$ are given by the relevant column vector of the matrix
$M^i$. The periodic approximant ``converges" slowly to the ISQL in
comparison with the case of binary QLs because
eq\verb/./(\ref{eqn:infmtr4}) has a term proportional to $\varepsilon^i$.

Let $Q$ and $Q'$ be two QLs in a single cycle and let $M$ and $M'$ be the
corresponding inflation matrices. Then, there exist two matrices $K$, $L
\in GL^+(3, \;{\bf Z})$ so that $M = KM'K^{-1}$ and $M' = LML^{-1}$; the
characteristic polynomial is common between $M$ and $M'$. We may say that
two inflation matrices satisfying this condition are modular equivalent to
each other. A necessary condition for two inflation matrices to be modular
equivalent is that both the scaling order and the signature are common
between the two.

The inflation matrix of a ternary QL is three-dimensional but is reducible
over ${\bf Q}$ because one of its eigenvalues is equal to 1 or $-$1. This
is consistent with that the dimension of the mother lattice of the QL is
not 3 but 2.

\subsection{Cyclic shift}\label{sec:cyclic shift}

If the three components, $S_x$ with $x = 1, \;\rho$ and $\omega$, of the
decoration vector, ${\bf S}$, of a 1DQL have commonly an interval $\alpha$
at the left ends of them, we can define a new string vector which is
written formally as ${\bf T}= \alpha^{-1}{\bf S}\alpha$; $T_x$ is derived
from $S_x$ by removing the interval $\alpha$ at the left end and,
subsequently, adding it at the right end. We shall call this procedure the
backward cyclic shift. By the assumption, $Q$ is expressed as an infinite
concatenation of $S_x$'s. Then, it is evident that $Q$ is expressed as an
infinite concatenation of $T_x$'s as well. Hence, ${\bf T}$ defines a new
DR and the relevant new subquasilattice, which is a shift of $Q'$ by
$\alpha$, while the subwindow for the new subquasilattice is a shift in
$E_\perp$ of the original one by $-{\tilde \alpha}$. The forward cyclic shift
is defined similarly. The two shift procedures can be repeated as long as the
conditions for the procedures are satisfied but will stop otherwise.
Starting from a DR, we can obtain in this way a full set of DRs. The full
set is a linearly ordered finite set. The DRs in the set have a common
decoration matrix, and are equivalent to one another because the full set
is retrieved from its any member.

The full set of DRs has a sort of mirror symmetry; a DR in the set and the
one at the mirror site are mutually mirror symmetric because each string of
the former is the mirror image of the relevant string of the latter. This
is a consequence of the mirror symmetry of the 1DQL. We shall call the
number of DRs in a full set multiplicity of the set. For example, the
multiplicity of the full set derived from the DR
eq\verb/./(\ref{eqn:mld21d}) is six. If multiplicity is odd, the central DR
must be symmetric; the relevant strings of the DR are palindromes. The full
set yields a full set of subquasilattices and that of the relevant
subwindows; the subquasilattices are translationally congruent with one
another. The subwindows have a common size and occupy different regions of
the original window $W$; different subwindows can partially overlap. The
configuration of the subwindows is symmetric in $W$, and the subwindow
associated with a symmetric DR (SDR) occupies a central region of $W$.

If two QLs which are combined by an SDR are similar geometrically, the SDR
gives an inflation rule, which we shall call the symmetric inflation rule
(SIR).

\section{Symmetric MLD}\label{sec:s-MLD}

\subsection{Symmetric MLD}\label{Symmetric MLD}

We may say generally that two QLs, $Q$ and $Q'$, which are SMLD is
symmetrically SMLD (s-SMLD) if local symmetry is common between them.
\cite{Ba99} We can define s-MLD (i.e., weak s-MLD) as well. If two QLs are
s-MLD, every special point of one of the two is a special point of the
other. However, the type of the special point can be different between the
two QLs; the four types of the special points are permuted in general
between the two QLs. The two QLs have common lattice points if and only if
the special point of the type $\Gamma$ is fixed on the permutation. We may
say that the relevant s-MLD relationship in this case is of the restrictive
type. A necessary condition for the two QLs to be
restrictively s-MLD is that the relevant two windows, $W$ and $W'$, is
concentric. \cite{Ni89a} If a QL $Q$ and its subquasilattice $Q'$ are
restrictively s-MLD, the center of each interval of $Q'$ is the
center of an interval of $Q$ because the center of an interval is a center
of a local symmetry of the type $X$, $Y$ or $C$.

We shall proceed to the nonrestrictive s-MLD relationship. Each simple DR in
\S \ref{sec:DR} is associated with a subquasilattice, $Q'$,
obtained by a decimation rule in which surviving lattice points are in the
R-enviroment of the form $*\bullet x$ with $x$ being a specified interval.
The relevant DR is apparently asymmetric. Remind here that the center of
each interval of type $x$ is a center of a local symmetry of the original
QL, $Q$. The set of all the centers of this type of intervals form a QL,
$Q^*$, which is a shift of $Q'$: $Q^* = Q' + \frac{1}{2}x$. The new QL,
$Q^*$, has no common lattice point with $Q$. Nevertheless it is MLD with
$Q$ because so is with $Q'$. We shall call $Q^*$ a pseudo-subquasilattice
of $Q$. It is important that $Q$ and $Q^*$ are s-MLD.
Pseudo-subquasilattices of $Q$ can be classified into three types
corresponding to the three types of intervals, 1, $\rho$ and $\omega$.

The mother lattice $\Lambda^*$ of $Q^*$ is a shift of $\Lambda$: $\Lambda^*
= \Lambda + {\bf x}/2$ with ${\bf x} = (x, \;-{\bar x}) \in \Lambda$. The
two mother lattices are not identical, and we may say that they are in a
symmetric arrangement. The s-MLD relationship between $Q$ and $Q^*$ is of
the nonrestrictive type. The four types of the special points of $\Lambda$
are transformed to those of $\Lambda^*$ but the types of the special points
are permuted on the transformation; $\Gamma$, $X$, $Y$ and $C$ are
transformed to $X$, $\Gamma$, $C$ and $Y$, respectively, if $x = 1$, for
example. This results in the same change of four types of the special
points between $Q$ and $Q^*$.

Let ${\bf S}$ (resp\verb/./ ${\bf S}^*$) be the decoration vector of a DR
combining $Q$ and $Q'$ (resp\verb/./ $Q^*$). Then, ${\bf S}^*$ is derived
from ${\bf S}$ by ``a cyclic half-shift" which is formally represented by
the equation ${\bf S}^* := (\frac{1}{2}x)^{-1}{\bf S}(\frac{1}{2}x)$. For
example, ${\bf S}^* := (\frac{1}{2}\rho\frac{1}{2},
\;\frac{1}{2}\omega^r\frac{1}{2}, \;\frac{1}{2}\omega^{r+1}\frac{1}{2})$ if
${\bf S} = (1\rho, \;1\omega^r, \;1\omega^{r+1})$ which is the one
associated with the DR B (see the first row of
eq\verb/./(\ref{eqn:mld20a})). Every component of ${\bf S}^*$ is symmetric
as it should be. We may call this type of DR a fractional DR, while the
conventional one an integral DR. A fractional DR is only used to represent
an SDR. \cite{ST86, Lu93} Fractional DRs can be classified into three types
as well. Every DR in \S \ref{sec:DR} is changed by a cyclic
half-shift into a fractional SDR. The restrictive s-MLD relationship does
not allow fractional SDRs.

A component of a fractional SDR is sandwiched by two half intervals but the
half intervals should be regarded to have polarities which are opposite
between the two because the half interval on the right or the left is the
first or the second half of the full interval. To see it we shall consider
the case where the relevant interval, $x$, is symmetrically decorated and
changed to a symmetric composite interval, $x'$. $x'$ can be divided evenly
into two parts which are the mirror images of each other. Each of the two
parts is composed only of full intervals if the number of intervals in $x'$
is even but it includes a half interval at the left or right end depending
on its polarity. In any event, a composite of two SDRs is also an SDR even
if either of the two or both are of the fractional type.

Let $Q_0 \succ Q_1 \succ Q_2 \succ {}\cdots$ with $Q_0 := Q$ be a
semi-infinite chain of subquasilattices generated by the SM. Then we can
obtain another semi-infinite chain of pseudo-subquasilattices, $Q^*_0 \succ
Q^*_1 \succ Q^*_2 \succ {}\cdots$ with $Q^*_0 := Q$. Let $Q'$ be one of the
member of the former chain and $Q^*$ be the corresponding member of the
latter chain. Then $Q^*$ is s-MLD with $Q$, and we can conclude that {\it
an MLD class is also an s-MLD class.}
There exists $\zeta \in {\bf Z}_{\rm H}[\tau]$ so that $Q^* = Q' + \zeta$.
$Q^*$ is a true subquasilattice of $Q$ if $\zeta$ is a $\tau$-integral but
a pseudo-subquasilattice otherwise. The DR combining $Q$ and $Q'$ in the
former case is equivalent to the SDR combining $Q$ and $Q^*$. In the latter
case, the fractional part of $\zeta$ is written as $\frac{1}{2}x$ with $x$
being equal to 1, $\rho$ or $\omega$. Then, the fractional SDR, ${\bf
S}^*$, combining $Q$ and $Q^*$ is of type $x$. It is changed by the inverse
cyclic-half-shift into an integral DR: ${\bf S}_x := (\frac{1}{2}x){\bf
S}^*(\frac{1}{2}x)^{-1}$. This DR is equivalent to the DR combining $Q$
and $Q'$. The decoration vector, ${\bf S}_x$, has the following properties:
i) its each component has the interval $x$ at its left end, and ii) the
three components of the vector, $x^{-1}{\bf S}_x$, are palindromes.
Two of the palindromes are odd-membered, and the central intervals
form with the interval $x$ the triplet $\{1, \;\rho, \;\omega\}$. This is derived
from the fact that special points are common between $Q$ and $Q^*$ on account of
the two QLs being s-MLD. We may call ${\bf S}_x$ a quasi-symmetric decoration vector.

A necessary and sufficient condition for a DR to
be equivalent to an integral SDR is that the decoration matrix $M$ is
congruent with a permutation matrix if it is considered in modulo 2. If
this condition is not satisfied, one of the three row vectors of $M \bmod
2$ is equal to $(1, \;1, \;1)$; the relevant row is the first, the second
or the third according as the type of the fractional DR is of the interval,
1, $\rho$ or $\omega$, respectively. A similar argument applies to the
inflation rule of an ISQL. If the inflation matrix is not congruent with a
permutation matrix in modulo 2, we have to repeat the IR at most four times
to obtain another IR which is equivalent to an integral SIR.

The three components of a decoration vector ${\bf S}_x$ are not
independent. To see it, we take a QL with a global mirror symmetry from the
relevant LI class, where the mirror symmetry is assumed to be of the same
type as that of the center of the interval $x$. We can cut the QL and
obtain a finite symmetrical segment, $X$, which may be of the fractional
type. If we divide $X$ evenly, and paste the original ends of the two
pieces, we obtain another string, which yields the same ``periodic
approximant" to the QL as that given by $X$. It is not difficult to show
that the three components of ${\bf S}_x$ are strings of this type. Since
the third component is the longest of the three, first two components are
in a sense parts of the third. A similar argument applies to the case of
the integral SDR.

We can distinguish the self-similarity in the MLD relationship and the one
in the s-MLD relationship; a QL, $Q$, is symmetrically self-similar if there
exists another QL which is not only s-MLD with $Q$ but also geometrically
similar to $Q$. From a conclusion presented above, we can conclude that
every QL (of type I or II) is symmetrically self-similar; the ratio of the
self-similarity is equal to that in the conventional self-similarity. Since
IR is a special DR, we can conclude that a QL with an inflation symmetry
has always SIR and the ratio of the inflation symmetry is equal to that in
the conventional inflation symmetry.

We shall close this subsection by investigating the boundary QLs,
$Q_{\rm b}$ and $Q'_{\rm b}$, whose windows are equal to 2 and $2\sigma$,
respectively. Since $\varphi(2) = W_{m-1}$, $Q_{\rm b}$ is combined with
the binary QL, $Q_{m-1}$, by the DR, $1' = 1\rho$ and $\omega' = 1\omega$,
which follows from eq\verb/./(\ref{eqn:mld17p}). Similarly, $Q'_{\rm b}$
for the case of $e = 1$ is combined with $Q_{m-2}$ by the DR $1' = \rho 1$
and $\omega' = \rho\omega$. The same DR combines $Q'_{\rm b}$ with $\tau =
\tau_{\rm G}$ with the Fibonacci lattice. The DR which combines a boundary
QL with a binary QL is equivalent to a fractional SDR.

\subsection{Series of inflation-symmetric ternary QLs}\label{sec:rinfsymmtern}

Let $W'$ be a subwindow of an R-division of the window of a ternary QL $Q
:= Q(W)$. Then, its size is equal to the first, second or third component
of the interval vector eq\verb/./(\ref{eqn:infmtr1}) if the relevant
R-environment is of the type $*\bullet x$ with $x$ being 1, $\rho$ or
$\omega$, respectively. There exist similar relations among three QLs, $Q$,
$Q' := Q(W')$ and $Q^* = Q' + \frac{1}{2}x$ to those mentioned in the
preceding subsection. For example, the DR combining $Q$ and $Q'$ is
quasi-symmetric. However, the DR may not be simple because $W'$ can take
arbitrarily small value or, equivalently, the intervals in $Q'$ can
be arbitrarily large. Simple DRs as those in \S \ref{sec:DR} are
obtained if $W' > \tau^{-1}$ but the results are not presented here. If
$W'$ is not included in the fundamental interval, we must rescale it. Then,
we obtain three types of SMs:
\begin{equation}
        \varphi_1(W) = {\rm Sc}(W - 1), \quad \varphi_\rho(W) = {\rm Sc}(W
- \sigma), \quad \varphi_\omega(W) = {\rm Sc}(1 + \sigma - W).
\label{eqn:sm1}
\end{equation}
Therefore, $\varphi_{\rm SM}$ is the same as $\varphi_1$ (resp\verb/./
$\varphi_\rho$) in the interval ${\rm A} \cup {\rm B}$ (resp\verb/./ C),
while $\varphi_{\rm NSM}$ is the same as $\varphi_\omega$ in the interval B
or D.

The present SMs have an infinite number of discontinuities because $W'$ can
take arbitrarily small value. As a consequence, they have infinite series
of fixed points and the corresponding ISQLs. Simplest cases are the fixed
points satisfying the equation $W = \tau(W - 1)$, $W = \tau(W - \sigma)$
or $W = \tau(1 + \sigma - W)$ because the scaling order of the relevant
ISQL is equal to 1. We shall denote the relevant ternary QL as $T^e_{k, x}$
with $e = \pm 1$ and $x$ being 1, $\rho$ or $\omega$, where $k$ stands for
the class number of $\sigma$ but $k = 1$ for $x = 1$. An explicit
expression for the fixed point is given by
\begin{equation}
        W^e_{k, 1} := \frac{\tau}{\tau - 1}, \quad W^e_{k, \rho} :=
\frac{\tau\sigma}{\tau - 1}, \quad {\rm or} \quad W^e_{k, \omega} :=
\frac{\tau(1 + \sigma)}{\tau + 1}, \label{eqn:sm2}
\end{equation}
respectively. The class number must satisfy the inequality $1 \le k \le
m-2$ for $W^e_{k, \rho}$ but $1 \le k \le m-1$ for $W^e_{k, \omega}$. It
can be readily shown that $W^-_{1, 1}$, $W^-_{k, \rho}$ or $W^e_{k,
\omega}$ is a fixed point of $\varphi_{\rm SM}$, $(\varphi_{\rm SM})^k$ or
$(\varphi_{\rm NSM})^k$, respectively. If $W = W^e_{k, \omega}$, we obtain
$W' := (\varphi_{\rm NSM})^{k-1}(W) = W - k + 1 \in {\rm C}$, so that $W =
\varphi_{\rm NSM}(W')$. It follows that $Q' := Q(W')$ is combined with $Q$
by the $k-1$ step cascade process, $(1, \;1^{k-1}\rho, \;1^{k-1}\omega)$,
and, conversely, $Q$ is combined with $Q'$ by the DR given by
eq\verb/./(\ref{eqn:nega2}) or (\ref{eqn:nega5}) according as $e = -1$ or
$e = 1$, respectively.\cite{FootNote10} The IRs of $Q$ and $Q'$ are given
as compositions of the two DRs. The case $k = 1$ is simplest because
$W^e_{1, \omega}$ is a fixed point of the $\varphi_{\rm NSM}$, and the
inflation vector of $T^-_{1, \omega}$ (resp\verb/./ $T^+_{1, \omega}$) is
given by $(\omega, \;\omega 1(\rho 1)^{m-2}, \;\omega 1(\rho 1)^{m-1})$
(resp\verb/./ $(\omega, \;\omega\rho(1\rho)^{m-3},
\;\omega\rho(1\rho)^{m-2})$). A similar argument applies to $T^-_{k,
\rho}$. These arguments are needed to be slightly extended in order to
apply them to the case of $T^+_{k, \rho}$. We conclude this subsection by
noting that the inflation vector of $T^+_{1, 1}$ takes the form: $(1\rho,
\;1\rho\omega^{m-3}\rho, \;1\rho\omega^{m-2}\rho)$.

\subsection{List of inflation vectors of the ISQLs}\label{sec:list infvctr}

In Table \ref{table2} are presented the results for ISQLs given as the
representatives of the MLD classes listed in Table \ref{table1}. The QLs of
types 1, $\rho$, and $\omega$ belong to the series of ISQLs investigated in
the preceding section.

A QL can have two nonequivalent IRs, which must have different signatures.
This is the case for two IRs in the row Nos. 2, 3, and 9 of the table or the
first IR and the last one in the row No.6 of the table. If the first IR
of No.3 in the table is repeated twice we obtain an inflation vector being
equivalent to $(\rho 1\omega 1, \;\rho 1\omega^2 1, \;\rho 1(\omega^2
1)^2)$, whose scaling order is 2. The same is true for the second IR. On
the other hand, the relevant QL has the integral SIR, $(\omega 1\omega,
\;\omega 1\rho 1\omega, \;\omega 1\rho 1\omega 1\rho 1\omega)$, whose
scaling order is 2. However, this inflation vector is not equivalent to the
former, which is equivalent to a fractional SIR. It can be shown that the
two inflation vectors become equivalent if both are repeated twice.

\begin{table}
\begin{center}
\begin{tabular}{rccc}
\hline
{No} & type $1/\rho$ & type $\omega$ & integral SIR  \\ \hline
1 & $*$ & $(\omega, \;\omega\rho, \;\omega\rho 1\rho)$ & $(\omega, \;\rho
1\rho, \;\rho 1\rho 1\rho)$  \\
2 & $(1\rho^2, \;1\rho\omega\rho, \;1\rho(\omega\rho)^2)$ & $(\omega,
\;\omega\rho 1\rho, \;\omega(\rho 1\rho)^2)$ & $*$  \\ \hline
3 & $(1\rho, \;1\omega, \;1\omega^2)$ & $(\omega, \;\omega 1, \;\omega
1\rho 1)$ & $*$  \\
4 & $(1\rho^2, \;1\rho^2\omega\rho^2, \;1\rho^2\omega\rho\omega\rho^2)$ &
$*$ & $*$  \\
5 & $*$ & $(\omega 1\rho 1, \;\omega 1\rho 1\rho 1, \;\omega (1\rho 1\rho
1)^2)$ & $*$  \\ \hline
6 & $(\rho 1, \;\rho 1\omega^21, \;\rho 1\omega^31)$ & $(\omega 1, \;\omega
1^2\rho 1^2, \;\omega 1^2(\rho 1^2)^2)$ & $(\omega, \;\omega 1\rho 1\omega,
\;\omega 1\rho 1\rho 1\omega)$  \\
7 & $(1\rho, \;1\omega^3, \;1\omega^4)$ & $*$ & $*$  \\
8 & $*$ & $(\omega 1, \;\omega 1(\rho 1)^2, \;\omega 1(\rho 1)^3)$ & $*$
\\ \hline
9 & $(1\rho, \;1\rho\omega\rho, \;1\rho\omega^2\rho)$ & $(\omega,
\;\omega\rho 1\rho, \;\omega\rho (1\rho)^2)$ & $*$  \\
10 & $*$ & $(\omega 1, \;\omega 1\rho 1, \;\omega (1\rho 1)^2)$ & $(\omega,
\;\omega\rho\omega, \;\omega\rho 1\rho\omega)$  \\
11 & $*$ & $(\omega 1^2, \;\omega 1, \;\omega 1^2\rho 1^2)$ & $*$  \\
\hline
\end{tabular}
\end{center}
\caption{Inflation vectors of the ISQLs of several type II QLs which are
listed in Table \ref{table1}. Those of the ISQLs with $m^e = 6^+$ ($\tau = 3 +
2\sqrt{2}\, $) are omitted to save the space.}
\label{table2}
\end{table}

\subsection{Restrictive s-MLD}\label{sec:restrictive s-MLD}

If the outermost two subwindows of an LR-division of a window are
deleted, we obtain a new subwindow $W'$ which is concentric with $W$. Its
size is given with the original window as $W' = 2\sigma - W$ if $W \in
\Gamma^{\rm c} \; (= A \cup B)$ but as $W' = 2 - W$ if $W \in \Gamma \; (=
C)$. By definition, $Q'$ is derived from $Q$ by decimating all the lattice
points in the local environment $1\bullet\rho$ or $\rho\bullet1$ in $Q$.
The two QLs are s-MLD because $W' \equiv -W \bmod 2{\bf Z}[\tau]$. More
strongly, we can show seperately for the four cases of the LR-divisions, A,
B, C and D, that the two QLs are related by an SDR, which can be obtained
by a similar argument to those in \S \ref{sec:DR}. We do not,
however, present the SDRs. We only mention that the signatures of all the
SDRs are minus because the relevant MLD-relationships are all improper.

We can define a saw map in connection with the SDRs above:
\begin{equation}
 \varphi_{\rm s}(W) = \left\{
     \begin{array}{ll}
        {\rm Sc}(2\sigma - W), & W \in \Gamma^{\rm c}, \\
        {\rm Sc}(2 - W), & W \in \Gamma,
     \end{array} \right. \label{eqn:sdr5}
\end{equation}
where $\sigma$ in the first row is regarded to be a function of $W$. A
fixed point of $\varphi_{\rm s}^{n}$ with $n$ being a natural number will
yield the window of an ISQL with an integral SIR. An example is given by
the first SIR in the last column in Table \ref{table2}.

Let $W'$ be the central subwindow of an LR-division of the window of a
ternary QL, $Q := Q(W)$. Then, $Q' := Q(W')$ is s-MLD with $Q$ because $W'
= |W - 2|$ if $W \in \Gamma^{\rm c}$ but $|W - 2\sigma |$ if $W \in
\Gamma$. The subquasilattice, $Q'$, is derived from $Q$ by retaining only
the lattice points in a symmetrical local environment of the type $x\bullet
x$ with $x$ being a specified interval. The two QLs are related by an
integral SDR, in which three strings representing the inflated intervals
are sandwiched by the interval $x$, where the sandwiched parts
do not include $x$. Since $W'$ can take arbitrarily small value, the
situation is similar to that appeared in \S \ref{sec:rinfsymmtern}.
We can obtain a series of ISQLs derived from the
relevant saw map, fixed points of which satisfy the equation $W = \tau^n|W
- 2|$ or $W = \tau^n|W - 2\sigma|$ with $n$ being a natural number. Two
inflation vectors of this type are found in the last column of Table \ref{table2}.

We shall investigate integral SDRs between type I QLs. Let $Q$ and $Q'$ be
a type I QL and its subquasilattice, restrictively, and assume that they are combined by
an integral SDR. Then, the relevant two windows, $W$ and $W'$, are located
concentricly. Since the distance between an end of $W$ and that of $W'$ is
equal to $|W \pm W'|/2$, two $\tau$-integrals, $W$ and $W'$, must have a
common parity because $|W \pm W'|/2$ is a $\tau$-integral by the
assumption. Conversely, if this condition is satisfied,
the two QLs are restrictively s-MLD. The SIRs of binary QLs have been
extensively investigated in ref\verb/./\citen{TNO01}).

We shall conclude this subsection by considering ISQLs of ternary QLs which
are of type I. Since the window of a binary QL is never divisable by 2, a
ternary QL cannot be combined by an integral SDR with a binary QL, if its
window is an even $\tau$-integral. The set of all the type I QLs whose
windows are even $\tau$-integrals form a subclass of the MLD class formed
by all the type I QLs. We can take the boundary QLs, $Q_{\rm b}$ and
$Q'_{\rm b}$, as representatives of the respective subclasses because their
windows are even $\tau$-integrals. The relevant boundary QL for $e = -1$ is
$Q'_{\rm b}$ (resp\verb/./ $Q_{\rm b}$) if $m = 1$ (resp\verb/./ $m \ge 2$).
The relevant window, $W = 2\tau^{-1}$ (resp\verb/./ $W = 2$), is a fixed
point of the SM, $\varphi_{\rm s}$, and the boundary QL has an SIR
represented by the inflation vector $(\rho, \;\omega, \;\rho 1\rho)$ with
$\rho = \tau$ (resp\verb/./ $(\omega, \;1(\rho 1)^{m-1}, \;1(\rho 1)^m)$
with $\rho = \tau - 1$). The situation is different for the case of $e = 1$
because the two windows of the two boundary QLs, $Q_{\rm b}$ and $Q'_{\rm
b}$, are given by $W_{\rm b} = 2$ and $W'_{\rm b} = 2 - 2\tau^{-1}$,
which form a 2-cycle of $\varphi_{\rm s}$. The SDR combining $Q_{\rm b}$ with $Q'_{\rm
b}$ is given by $(1, \;\omega, \;1\rho 1)$ with $\rho = \tau - 2$, while
the inverse SDR is by $(\omega, \;\rho(1\rho)^{m-3}, \;\rho(1\rho)^{m-2})$
with $\rho = \tau - 1$. It follows that the two boundary QLs have integral
SIRs $(1\rho 1, \;\omega(1\omega)^{m-3}, \;\omega(1\omega)^{m-2})$ and
$(\omega, \;\rho(1\rho)^{m-2}, \;\omega\rho(1\rho)^{m-3}\omega)$,
respectively.

\section{MLD Relationship in the Section Method}\label{section method}

The section method of constructing a QL from a mother lattice is equivalent
to the cut-and-projection method but it is illuminating to see the MLD
relationship from the point of view of the section method. \cite{Ya96} In
this method ``the atomic surface" is located on every lattice point of the
mother lattice, and the QL is given as a section of the hyper-crystal of
the atomic surfaces through the physical space, where the size of the
atomic surface is equal to the size of the window in the cut-and-projection
method. It is evident that a hyper-crystal has is its own LI-class of QLs.
We shall denote by $\Lambda\{W\}$ the hyper-crystal specified by the mother
lattice $\Lambda$ and the window (atomic surface) with size $W$. The
lattice points of a QL are regarded to be atoms in the section method. We
consider that the physical space $E_\parallel$ is horizontal, while the
internal space $E_\perp$ vertical. Then the two ends of an atomic surface
can be distinguished by their vertical level. The so-called gluing
condition for the hyper-crystal consists of the two conditions, i) the
front end of every atomic surface has its partner which is the rear end of
another atomic surface and ii) the two partnered ends are in the same
level. If the physical space $E_\parallel$ is slightly shifted along the
internal space, a part of the atoms of the QL disappear but new atoms
appear instead. Then, the atoms are conserved locally if the hyper-crystal
satisfies the gluing condition; disappearence of an atom at a site and
appearence of a new atom at another site works together and the distance
between the two site is equal to the distance between the two partnered
ends. \cite{Fr86, Ka89} The condition is always satisfied by type I QLs
(hyper-crystals) but not by type II QLs. We shall call the present gluing
condition the self-gluing condition in order to distinguish it from a
gluing condition to appear below.

Exactly speaking, the MLD relationship is not a relationship between two
QLs but between two LI-classes. Therefore, it is a relationship between two
hyper-crystals. We begin with the improper MLD relationship, where the two
relevant windows satisfy the condition, $\lambda := W + W' \in {\bf
Z}[\tau]$, which is rewritten as
\begin{equation}
 \frac{W}{2} = \frac{\lambda}{2} - \frac{W'}{2} \;{\rm or}\; -\frac{W}{2} =
-\frac{\lambda}{2} + \frac{W'}{2}. \label{eqn:section1}
\end{equation}
Let ${\bf x} \in \Lambda$ be a lattice vector which is projected onto
$\lambda$. Then, these equations can be interpreted as an inter-gluing
condition between two hyper-crystals $\Lambda\{W\}$ and $\Lambda'\{W'\}$
with $\Lambda' := {\bf x}/2 + \Lambda$. Thus, the condition for the
 improper MLD relationship between two hyper-crystals is equivalent to the
inter-gluing condition between the two. Since ${\bf x} \in \Lambda$, every
lattice point of $\Lambda'$ is a special point of $\Lambda$; $\Lambda'$ is
a lattice of all the special points whose type is the same as that of ${\bf
x}/2$. It is evident that two QLs obtained from the two hyper-crystals,
$\Lambda\{W\}$ and $\Lambda'\{W'\}$, by cutting them by a common physical
space are s-SMLD. We may say that two windows $W$ and $W'$ are
complementary to each other if they satisfy the condition, $W + W' \in {\bf
Z}[\tau]$. If $\Lambda\{W\}$ is a type II hyper-crystal, so is
$\Lambda'\{W'\}$.

The union of two hyper-crystals above yields a non-Bravais type
hyper-crystal, $\Lambda\{W\} \cup \Lambda'\{W'\}$, provided that ${\bf x}/2
\notin \Lambda$; the relevant non-Bravais type QLs have two types of atoms
derived from the two types of atomic surfaces. This non-Bravais type
hyper-crystal satisfies a sort of the gluing condition even if the two component
hyper-crystals are of the type II. Note, however, that the local
conservation of atoms holds only when the two types of atoms are not
distinguished. Therefore, the present gluing condition may be called the
heterogeneous gluing condition.

The condition for the proper MLD relationship is equivalent to $W/2 =
\lambda/2 + W'/2$ with $\lambda \in {\bf Z}[\tau]$.
This relationship has a simple meaning in the
geometrical relation between the two hyper-crystals $\Lambda\{W\}$ and
$\Lambda'\{W'\}$ but we shall not persuit it.

A QL belonging to an MLD class of type I QLs is often caused breaking of
one or more types of local symmetries. \cite{Ni89d} The symmetry breaking
is a consequence of a mismatch between the gluing condition and the
symmetry. It never occurs for the case of a type II QL.

\section{Summary and Further Remarks}

A classification of 1DQLs into mutual local-derivability (MLD) classes has
been completed for each $\tau$ on the basis of geometrical and
number-theoretical considerations; a generic 1DQL is ternary. All the type
I QLs, whose windows are $\tau$-integrals, form one MLD class
but there exist an infinite number of MLD classes of
type II QLs, whose windows are $\tau$-rationals. For every MLD class, we
can choose an inflation-symmetric member as its representative, and a
non-inflation-symmetric member is given as a decoration of the
representative.
An algorithm for deriving an inflation-symmetric subquasilattice of a given
QL is presented. Also, several properties of a number of important MLD
classes of type II QLs are investigated. Symmetric decoration rules and
symmetric inflation rules are investigated extensively.

One of the eigenvalues of the inflation matrix of a ternary QL is equal to
1 or $-$1, while the absolute value of another non-Frobenius eigenvalue is
less than 1, so that the inflation matrix is marginal in a classification
scheme for ``IRs". \cite{Bo86} This may be the origin of an anomalous
behavior of the energy spectrum of the one-electron Hamiltonian on a type
II QL. \cite{FuNi00, FuNi01} A slow convergence of the periodic approximant
as mentioned in \S \ref{sec:inf} is also a feature of a quasiperiodic
structure generated by a marginal inflation rule. \cite{Bo86, Au88}

The structure factor of a 1DQL is composed of Dirac peaks which are indexed
by the Fourier module of the QL. It is known that the Fourier module is
determined solely by the mother lattice, so that the size of the window is
indifferent to it. Therefore, the Fourier module is incapable of
discriminating two QLs which belong to different MLD classes. It should be
emphasized in this respect that the Fourier module of an ISQL is not
determined solely by the ratio of the inflation-symmetry. For example, a
type II QL derived from the mother lattice characterized by $\tau_{\rm G}$
has $2 + \sqrt{5}\, $ as its ratio of the inflation-symmetry but its Fourier
module is essentially different from that of a QL derived from the mother
lattice characterized by $2 + \sqrt{5}\, $ and a similar comment applies to
a QL derived from the mother lattice characterized by $\tau_{\rm S}$ if its
ratio of the inflation-symmetry is equal to $(\tau_{\rm S})^2 = 3 +
2\sqrt{2}\, $.

There exists a different but interesting view for type II QLs. This is
presented in Appendix \ref{DifferentView}.

It is known that there exists a close relation between 1DQLs and QLs in
higher dimensions. \cite{DuKa85, ST86, Lu87, Lu93, HNL94} It can be shown
that a Bravais type QL in $d$-dimension is constructed by the use of a
star-like arrangement of 1DQLs. Therefore, the present theory is,
basically, applicable to a classification of higher-dimensional QLs into
MLD classes. Since higher-dimensional QLs have a high point symmetry, it is
important to classify them by means of the s-MLD which is developed in
\S \ref{sec:s-MLD}. A full presentation of the results for
higher-dimensional QLs will be made in a separate paper. \cite{Ni01}

Ternary IRs were introduced by S. Aubry et al \cite{Au88} for circle
sequences (CSs), which are generated by circle maps characterized by
quadratic fields. They are similar to the ternary IRs in this paper
although a CS is not a geometrical object but an infinite sequence of two
digits ``0" and ``1". Quite recently, the authors have succeeded in
revealing a general rule which relates a 1DQL to a CS. However, we do not
present here the general rule and its proof because a considerable space is
required (for a preliminary argument, see ref\verb/./\citen{FuNi01})).

It is known that an aperiodic structure generated by ``an IR" is not
necessarily given by the projection method with a simple window.
\cite{LuGoJa93, Au88} ISQLs presented in this paper are, to our best
knowledge, first examples of ternary structures which are not only
generated by IRs, but also given by the projection method with non-fractal
windows. Note, however, that most ternary 1DQL of type I or II does not
obey any IR though it is given as a decoraton of an ISQL.

A similar residue class to the one defined with respect to window sizes
appears in ``torus parametrization" of phases of QLs. \cite{Ba97} The SM in
the present theory induces a permutation among the residue classes in a
scaled residue class, and a similar permutation appears when the inflation
symmetry of a binary QL is treated in the theory of the torus parametrization.
Consequently, a part of mathematics are common between the two theories. We
should mention, however, that the window and the phase play entirely
different roles from each other on determining the structure of a QL.

\bigskip
\noindent
{\bf Acknowledgments}:
\bigskip

A part of this work was done while N. F. was staying at the Institute
of Crystallography, Russian Academy of Sciences (Moscow), under the grant:
Obuchi Fellowship Program. He is most grateful to V. E. Dmitrienko and the
members of the Laboratory of Theoretical Studies for their kind
hospitality. Financial support from the Japan-Russia Youth Exchange Center
is acknowledged.

\appendix

\section{Number Theory of Quadratic Fields}\label{NumberTheory}

Several results in the number theory of quadratic fields are summarized here
within a scope necessary in this paper. Some of them can be found, for
example, in ref\verb/./ \citen{Ha79}) but others are our originals.

\subsection{General}
The quadratic field ${\bf Q}[\tau] := \{r + s\tau \, |\, r, s \in {\bf Q}\}$
is a dense subset of ${\bf R}$, the real field, while the module ${\bf
Z}[\tau]$ is a dense subset
of ${\bf Q}[\tau]$: ${\bf Z}[\tau] \subset {\bf Q}[\tau] \subset {\bf R}$.
Of the three sets, ${\bf Z}[\tau]$ and ${\bf Q}[\tau]$ are countable but
${\bf R}$ is uncountable. We shall call a real number a $\tau$-integral,
$\tau$-rational or $\tau$-irrational according as it belongs to ${\bf
Z}[\tau]$, ${\bf Q}[\tau] - {\bf Z}[\tau]$ or ${\bf R} - {\bf Q}[\tau]$,
respectively. However, $\tau$-integrals are sometimes regarded as special
$\tau$-rationals.

The golden ratio satisfies the equalities, $\tau_{\rm G}^2 = \tau_{\rm G} +
1$ and $\tau_{\rm G}^3 = 2 + \sqrt{5}\, $, and we obtain ${\bf Q}[\tau_{\rm
G}] = {\bf Q}[(\tau_{\rm G})^2] = {\bf Q}[2 + \sqrt{5}\, ]$. On the other
hand, ${\bf Z}[\tau_{\rm G}] = {\bf Z}[(\tau_{\rm G})^2] \ne {\bf Z}[2 +
\sqrt{5}\, ]$; the module ${\bf Z}[\tau]$ with $\tau = 2 + \sqrt{5}\, $
is not closed as an integral domain because $\tau_{\rm G}$ is
not a $\tau$-integral in ${\bf Z}[\tau]$. A similar relation holds between
$\tau_{\rm S} := 1 + \sqrt{2}\, $ (the silver ratio) and $\tau = 3 +
2\sqrt{2}\, $ which satisfies $\tau = \tau_{\rm S}^2$.

Let ${\bf R}^+$ be the set of all the positive real numbers and $X$ be a
subset of ${\bf R}$. Then, the symbol $X^+$ stands for $X \cap {\bf R}^+$
throughout this paper.

If a $\tau$-integral is factorized into several $\tau$-integrals, each
factor is called a divisor of the $\tau$-integral. Since $\tau, \:\tau^{-1}
\in {\bf Z}^+[\tau]$, $\tau^n$ with $n \in {\bf Z}$ is always a divisor of
every $\tau$-integral. Therefore, a number of the form $\tau^n$ is a trivial
divisor and is not considered usually to be a divisor. If two positive
$\tau$-integrals $\mu$ and $\mu'$ satisfy $\mu/\mu' = \tau^n$ with $n \in
{\bf Z}$, we say that $\mu$ and $\mu'$ are companions of each other, and
denote as $\mu \cong \mu'$. A number and its companions are not
distinguished in a consideration of multiplicative properties of ${\bf
Z}^+[\tau]$. A $\tau$-integral is called a prime number if it has no
divisors but itself. A composite number can be ``uniquely" factorized into
primes for some $\tau$'s, e.g., $\tau_{\rm G}$, $\tau_{\rm S}$, $2 +
\sqrt{3}\, $, etc., but not for $2 + \sqrt{5}\, $ and $3 + 2\sqrt{2}\, $, for
example.
The norm of a $\tau$-integral $\mu \;(\ne 0)$ is defined by $N(\mu) :=
|\mu{\bar \mu}|$, which is a natural number. A sufficient condition for
$\mu \in {\bf Z}^+ [\tau]$ to be a prime is that $N(\mu)$ is a prime in
${\bf Z}^+$.

If $\zeta \in {\bf Q}^+[\tau]$, it is represented as a simple fraction:
$\zeta = \lambda/\mu$ with $\lambda$, $\mu \in {\bf Z}^+[\tau]$; $\lambda$
and $\mu$ are coprime, namely, they have no common divisors.

The set of all the unimodular matrices in $d$-dimensions form an infinite
group, $GL(d, \;{\bf Z})$. The set composed of all the Frobenius matrices
in $GL(d, \;{\bf Z})$ is denoted as $GL^+(d, \;{\bf Z})$, where a Frobenius
matrix (exactly, Perron-Frobenius matrix) is a matrix whose elements are
all nonnegative. Note that $GL^+(d, \;{\bf Z})$ is not a group but a
semigroup because the inverse of a matrix in $GL^+(d, \;{\bf Z})$ does not
belong to $GL^+(d, \;{\bf Z})$ unless it is a permutation matrix.

\subsection{Residue classes and scaled residue classes}\label{scaled
residue class}

Let $\mu \not\cong 1$ be a positive $\tau-$integral. Then, two
$\tau-$integrals $\lambda$ and $\lambda'$ are defined to be congruent in
modulo $\mu$ with each other if $\mu$ divides $\lambda - \lambda'$. The
congruence relation is written as $\lambda \equiv \lambda' \bmod \mu$.
Using the relation, we can divide the set ${\bf Z}[\tau]$ into residue
classes, whose number is equal to $N(\mu)$; the residue class to which
$\lambda \in {\bf Z}[\tau]$ belongs is given by $\lambda + \mu{\bf
Z}[\tau]$, which is denoted as $\lambda \bmod \mu$. A residue class is
called irreducible if a number belonging to the class is coprime
with $\mu$. Let ${\cal I}_{\mu}$ be the set of all the irreducible residue
classes in modulo $\mu$. Then the number of the elements in the set is
given by the Eulerian function $\Phi(\mu)$. If $\mu \cong \mu'$, then
$\Phi(\mu) = \Phi(\mu')$. In particular, $\Phi(\mu) = N(\mu) - 1$ if $\mu$
is prime in ${\bf Z}[\tau]$. If $\lambda$ and $\mu$ are coprime,
$\Phi(\lambda\mu) = \Phi(\lambda) \Phi(\mu)$.

A multiplication of $\tau$ onto ${\bf Z}[\tau]$ induces a linear
automorphism of ${\bf Z}[\tau]$, so that the same is true for its effect on
${\cal I}_{\mu}$. The automorphism is a permutation over the finite set
${\cal I}_{\mu}$. \cite{Ni89b} The order $q$ of the permutation is equal to
the smallest natural number so that $\mu$ is a divisor of $\tau^q - 1$. The
number $q$ is called the index of $\mu$. The set ${\cal I}_{\mu}$ is
divided by the permutation into orbits; the number of the orbits is given
by $q^{-1}\Phi(\mu)$, so that $q$ is a divisor of $\Phi(\mu)$. The orbit
including $\lambda \bmod \mu$ is given by $\{\lambda \bmod \mu,
\;\tau\lambda \bmod \mu, \;\cdots, \;\tau^{q - 1}\lambda \bmod \mu\}$. The
number of the orbits is one only if $q = \Phi(\mu)$.

The residue classes discussed above are considered in ${\bf Z}[\tau]$ but
it is convenient to consider residue classes in ${\bf Q}[\tau]$; the
residue class to which $\zeta \in {\bf Q}[\tau]$ belongs is the set of
numbers defined by $\zeta + {\bf Z}[\tau]$; a necessary and sufficient
condition for $\zeta$ and $\zeta'$ in ${\bf Q}[\tau]$ to belong to a common
residue class is $\zeta \equiv \zeta' \bmod {\bf Z}[\tau]$.
If $\zeta = \lambda/\mu \in
{\bf Q}^+[\tau]$, this residue class is derived from the residue class
$\lambda \bmod \mu$ by dividing it by $\mu$. We define here
the index $q$ of $\zeta$ by that of its denominator $\mu$
in ${\bf Z}[\tau]$; $q$ is the smallest natural number satisfying the
condition $(\tau^q - 1)\zeta \in {\bf Z}[\tau]$. Remember that the index is
independent of the numerator $\lambda$ of $\zeta$ and that $\zeta$ and
$\tau^n\zeta$ with any $n \in {\bf Z}$ have a common index. The index of a
$\tau$-integral, in particular, is defined to be 1. It is evident that the
division of ${\bf Z}[\tau]$ into residue classes with respect to $\mu \in
{\bf Z}^+[\tau]$ is equivalent to that of $\mu^{-1}{\bf Z}[\tau]$ into
residue classes with respect to ${\bf Q}[\tau]$.

Let us define the scaled residue class for $\zeta \in {\bf Q}[\tau]$ by
\begin{equation}
\Sigma(\zeta) := (\zeta + {\bf Z}[\tau])\cup(\tau \zeta + {\bf
Z}[\tau])\cup\;\cdots \;\cup(\tau^{q - 1}\zeta + {\bf Z}[\tau])
\label{eqn:nt1}
\end{equation}
with $q$ being the index of $\zeta$. Then, $\Sigma(\zeta) + \nu =
\Sigma(\zeta)$ for ${}^\forall \nu \in {\bf Z}[\tau]$ and $\tau
\Sigma(\zeta) = \Sigma(\zeta)$. Let $\zeta, \;\zeta' \in {\bf Q}[\tau]$.
Then, $\Sigma(\zeta) = \Sigma(\zeta')$ or $\Sigma(\zeta) \cap
\Sigma(\zeta') = \phi$ according as there exists $n \in {\bf Z}$ so that
$\zeta \equiv \tau^n\zeta' \bmod {\bf Z}[\tau]$ or there does not,
respectively. We shall call, incidentally, this relationship between two
numbers a scaled-congruence relationship in modulo ${\bf Z}[\tau]$.

The module ${\bf Z}[\tau]$ acts on ${\bf Q}[\tau]$ as a translation group.
A scaled residue class is nothing but an orbit in ${\bf Q}[\tau]$ with
respect to the translation-scaling group which is obtained from the
translation group by adding the scaling operation to it. The scaling by
$\tau$ induces a cyclic permutation among the members in the right-hand
side of (\ref{eqn:nt1}).

The index $q$ is common for all the $\tau$-rationals in a scaled residue
class; it is a proper number of the class. The scaled residue class is
a subset of ${(\tau^q - 1)}^{-1}{\bf Z}[\tau]$, which can be divided into a
finite number of residue classes; the number is equal to $N(\tau^q - 1)$.
Therefore, the number of scaled residue classes whose indices are specified
to be $q$ does not exceed $q^{-1}N(\tau^q - 1)$, which is equal to $m - 1 -
e$ if $q = 1$.

\subsection{Generalized parities of $\tau$-integrals}\label{generalized parity}

Let $\tau^2 - m\tau + e = 0$ be the defining equation for $\tau$. Then,
$\tau^2 + \tau + 1 \equiv 0 \bmod 2$ if $m$ is odd but $(\tau \pm 1)^2
\equiv \tau^2 + 1 \equiv 0 \bmod 2$ otherwise. Let $\lambda$ be a
$\tau$-integral written as $\tau^q$ with $q$ being a natural number. Then,
we can show readily that $\lambda^2 + \lambda + 1 \equiv 0 \bmod 2$ if both
$m$ and $q$ are odd and, besides, $q$ is not a multiple of 3. Otherwise, we
obtain $\lambda^2 + 1 \equiv 0 \bmod 2$. In the former case, $\lambda \pm
1$ are never divisable by 2 but $\lambda^3 \pm 1$ are. In the latter case,
$\lambda \pm 1$ may not be divisable by 2 but $\lambda^2 \pm 1$ are.
A $\tau$-integral is called even if it is divisable by 2. A $\tau$-integral
being not even is called semi-even or odd according as its square is
divisable by 2 or not, respectively. For example, $1 + \tau$ is semi-even
if $m$ is even. The generalized parities are invariant against scaling by
$\tau$.

We may write
\begin{equation}
{\bf Z}[\tau] = 2{\bf Z}[\tau] \cup (1 + 2{\bf
Z}[\tau])\cup(\tau + 2{\bf Z}[\tau])\cup(1 + \tau + 2{\bf Z}[\tau]),
\end{equation}
which represents a division of ${\bf Z}[\tau]$ in modulo 2. The first of the four
residue classes is composed of all the even $\tau$-integrals but the next
two are only composed of the odd $\tau$-integrals. On the other hand, the
last one is composed of all the semi-even $\tau$-integrals if $m$ is even
but of the odd $\tau$-integrals otherwise. The four residue
classes yield similar four residue classes into which the set ${\bf Z}_{\rm
H}[\tau] := \frac{1}{2}{\bf Z}[\tau]$ is divided:
\begin{equation}
{\bf Z}_{\rm H}[\tau] = {\bf Z}[\tau] \cup\left(\frac{1}{2} + {\bf
Z}[\tau]\right)\cup\left(\frac{\tau}{2} + {\bf
Z}[\tau]\right)\cup\left(\frac{1 + \tau}{2} + {\bf Z}[\tau]\right).
\label{eqn:nt2}
\end{equation}
We shall represent the four classes by $\Gamma$, $X$, $Y$ and $C$,
respectively. The four residue classes are permuted among themselves on the
scaling by $\tau$ because ${\bf Z}_{\rm H}[\tau]$ as well as ${\bf
Z}[\tau]$ is invariant against the scaling. Of the four residue classes,
$\Gamma$ is always invariant but others are not necessary so; $C$ is
invariant but $X$ and $Y$ are interchanged if $m$ is even, while $X$, $Y$
and $C$ are permuted cyclically in this order for an odd $m$. It follows
that ${\bf Z}_{\rm H}[\tau]$ is divided into three scaled residue classes,
${\bf Z}[\tau]$, $\Sigma(\frac{1}{2}) \;(=X \cup Y)$ and $C$, if $m$ is
even but into two scaled residue classes, ${\bf Z}[\tau]$ and
$\Sigma(\frac{1}{2}) \;(=X \cup Y \cup C)$, otherwise. Let $\zeta$ be a
$\tau$-rational belonging to ${\bf Z}_{\rm H}[\tau]$. Then it can be
readily shown that the index of $\zeta$ is equal to 3 for an odd $m$ but to
1 (resp\verb/./ 2) for an even $m$ if $\zeta$ belongs to $C$ (resp\verb/./
$\Sigma(\frac{1}{2})$).

\subsection{Symmetrized scaled residue classes}\label{Symmetrized scaled
residue classes}

It is obvious that $\Sigma(-\zeta) = -\Sigma(\zeta)$. $\Sigma(\zeta)$ has
the inversion symmetry if $\Sigma(\zeta) = -\Sigma(\zeta)$. For example,
the three residue classes $X$, $Y$ and $C$ are symmetric. A
$\tau$-rational $\zeta$ is classified into one of the three types: a) it
belongs to ${\bf Z}_{\rm H}[\tau]$, b) it does not but $\Sigma(\zeta) =
-\Sigma(\zeta)$ and c) $\Sigma(\zeta) \ne -\Sigma(\zeta)$. Let $q$ be the
index of a $\tau$-rational of type a. Then the equalities $(\tau^q \pm
1)\zeta \in {\bf Z}[\tau]$ hold for both signs. On the other hand, if
$\zeta$ is of the type b, we can show readily that its index $q$ is even
and, moreover, $(\tau^{q'} + 1)\zeta \in {\bf Z}[\tau]$ with $q' := q/2$,
which we shall call a half-index of $\zeta$; the equality $\tau^q - 1 =
(\tau^{q'} - 1)(\tau^{q'} + 1)$ is used on a proof of this statement. A
$\tau$-rational must be of the type a or b if its denominator is a divisor
of $\tau^{n} + 1$ for a non-negative integer $n$. A $\tau$-rational with
an odd index must be of type a or c. All the members of a scaled residue
class are of a common type.

There exist merely two or three scaled residue classes of type a for a
given $\tau$. Scaled residue classes of type b are minority in comparison
with those of type c. A scaled residue class whose half-index is $q'$ is
included in ${(\tau^{q'} + 1)}^{-1}{\bf Z}[\tau]$, and the number of such
scaled residue classes does not exceed ${q'}^{-1}N(\tau^{q'} + 1)$, which
is equal to $m + 1 + e$ if $q' = 1$.

A symmetrized scaled residue class (SSRC) is defined for $\zeta \in {\bf
Q}[\tau]$ by $\Sigma_{\rm sym.}(\zeta) := \Sigma(\zeta) \cup \Sigma(-\zeta)$, which is
identical to $\Sigma(\zeta)$ if $\zeta$ is of type a or b. An SSRC is
nothing but an orbit in ${\bf Q}[\tau]$ with respect to a group which is
obtained from the translation-scaling group by adding the inversion
operation to it. An SSRC has its proper index because the index is common
between $\Sigma(\zeta)$ and $\Sigma(-\zeta)$. SSRCs are naturally
classified into one of the three types, a, b and c according to the types
of the relevant $\tau$-rationals.

\subsection{Projection method}\label{Projection method}
There exists a bijection between the mother lattice $\Lambda$ and its
projection $\Lambda_\parallel = {\bf Z}[\tau]$:
\begin{equation}
\Lambda = \{(\nu, \;{\tilde \nu}) \, |\, \nu \in {\bf Z}[\tau] \} \label{eqn:nt3}
\end{equation}
with ${\tilde \nu} := -{\bar \nu}$. That is, ${\bf Z}[\tau]$ can be lifted
to $\Lambda$. Similarly, there exists a bijection between ${\bf Q}[\tau]$
and a set in the 2D space; the set is nothing but the one composed of all
the rational points of the 2D space with respect to $\Lambda$. To be more
specific, $\zeta \in {\bf Q}[\tau]$ can be lifted to $(\zeta, \;{\tilde
\zeta})$ with ${\tilde \zeta} := -{\bar \zeta}$. It follows that a
number-theoretical property of ${\bf Q}[\tau]$ is translated to a
geometrical property of the relevant set in 2D and vice versa. For example,
a multiplication of $\tau$ onto $\zeta \in {\bf Q}[\tau]$ is translated to
the hyperscaling operation on ${\bf x} := (\zeta, \;{\tilde \zeta})$.
For instance, eq\verb/./(\ref{eqn:nt1}) reads
\begin{equation}
\Lambda({\bf x}) := ({\bf x} + \Lambda)\cup(T{\bf x} + \Lambda)\cup\;\cdots
\;\cup(T^{q - 1}{\bf x} + \Lambda),
\label{eqn:nt4}
\end{equation}
where $\Lambda({\bf x})$ is generally a non-Bravais type lattice, which is derived
from $\Lambda$ by decorating it. This non-Bravais type lattice has the
hyperscaling symmetry.

It is known that the mother lattice $\Lambda$ in the projection method is
divided into several equivalent sublattices which are ``similar" to the
original lattice. \cite{Ni89b, Ni89c} By the correspondence between
$\Lambda$ and $\Lambda_\parallel = {\bf Z}[\tau]$, the sublattices are
related to residue classes of ${\bf Z}[\tau]$ with respect to an ideal of
${\bf Z}[\tau]$. The ideal is represented with $\mu \in {\bf Z}^+[\tau]$ as
$\mu{\bf Z}[\tau]$, and the number of the sublattices is equal to the norm
$N(\mu)$. Let $q$ be the index of $\mu$. Then, every residue class is
invariant against a multiplication of $\tau^q$. Therefore, these
sublattices are invariant against $T^q$ with $T$ being the hyperscaling.

\section{Different View for Type II QLs}\label{DifferentView}

The sublattices of $\Lambda$ discussed in Appendix \ref{Projection method}
naturally divide a QL, $Q$, derived from $\Lambda$ into different QLs,
which are mutually LI but not MLD with $Q$: $Q = Q_0 \cup Q_1 \cup \cdots
\cup Q_{n - 1}$ with $n=N(\mu)$. Let us assume that $Q$ is of type I. Then,
the window is given by a number, $\lambda \in {\bf Z}^+[\tau]$. It can be
shown readily that $Q_i^{\ast}:=\mu^{-1}Q_i$ is obtained by the projection
method from $\Lambda$ by using a window whose size is equal to
$\lambda/|{\tilde \mu}|$. Hence, $Q_i^{\ast}$'s are of type II provided
that $\lambda$ and ${\tilde \mu}$ are coprime. Conversely, every type II
QL is derived in this way from a type I QL. For example, the binary QL
$Q_1$ with $\tau = 1 + \sqrt{2}\, $ is divided into two type II QLs which
are locally isomorphic to No.3 in Table \ref{table1}; this is proven by taking $\mu =
\sqrt{2}\, $.


\begin{thebibliography}{50}
\bibitem{FootNote2} In this paper, we shall identify an infinite set of
numbers with an infinite-dimensional vector. Then, an affine transformation
of the set $W$ is defined by $a + b\, W := \{a + b\, x \, |\, x \in W$\},
where the symbol ``:=" means that the left hand side is defined by the
right hand side.
\bibitem{FootNote3} The mother lattice $\Lambda$ presented here turns to a
square lattice if $E_\perp$ is scaled up by $\tau$.
\bibitem{FootNote4} Actually, $F$ is not divided when $m=1$. The case $m=1$
will be sometimes exceptional in later arguments but we will not add notes
like this if obvious.
\bibitem{FootNote9} The abbreviation ``MLD" will be used not only for
``mutual local-derivability" but also for ``mutually locally-derivable".
\bibitem{FootNote8} Decoration is the inverse procedure of decimation, and
the decoration rule can be called, alternatively, by a decimation rule.
\bibitem{FootNote5} This follows from that the frequency vector $w$ assumes
 the form as given by eq\verb/./(\ref{eqn:infmtr1}) in \S \ref{infmtrx} .
\bibitem{FootNote10} This does not apply to $T^+_{m-1, \omega}$, whose
inflation vector is directly obtained as $(\omega 1^{m-2}, \;\omega
1^{m-3}, \;\omega 1^{m-2}\rho 1^{m-2})$.
\end{thebibliography}
\end{document}